\documentclass[sigconf,nonacm]{acmart}

\AtBeginDocument{%
  }

\usepackage{xcolor}
\usepackage{multirow}
\usepackage{colortbl}
\usepackage{soul}
\usepackage[subtle]{savetrees}
\usepackage{tcolorbox}
\tcbuselibrary{skins}
\newif\ifcomment
\commentfalse
\ifcomment
\newcommand{\sayak}[1]{{\bf \textcolor{blue}{Sayak: #1}}}
\newcommand{\shirin}[1]{{\bf \textcolor{purple}{Shirin: #1}}}

\else
\newcommand{\shirin}[1]{}
\newcommand{\sayak}[1]{}
\fi
\begin{document}


\title{"Explain, Don’t Just Warn!" - A Real-Time Framework for Generating Phishing Warnings with Contextual Cues}

\author{Sayak Saha Roy}
\affiliation{%
  \institution{The University of Texas at Arlington}
  \city{Arlington}
  \state{Texas}
  \country{USA}
}
\email{sayak.saharoy@mavs.uta.edu}

\author{Cesar Torres}
\affiliation{%
  \institution{The University of Texas at Arlington}
  \city{Arlington}
  \state{Texas}
  \country{USA}
}
\email{cearto@uta.edu}

\author{Shirin Nilizadeh}
\affiliation{%
  \institution{The University of Texas at Arlington}
  \city{Arlington}
  \state{Texas}
  \country{USA}
}
\email{shirin.nilizadeh@uta.edu}

\begin{abstract}
Anti-phishing tools typically display generic warnings that offer users limited explanation on why a website is considered malicious, which can prevent end-users from developing the mental models needed to recognize phishing cues on their own. This becomes especially problematic when these tools inevitably fail - particularly against evasive threats, and users are found to be ill-equipped to identify and avoid them independently.
To address these limitations, we present PhishXplain (PXP), a real-time explainable phishing warning system designed to augment existing detection mechanisms. PXP empowers users by clearly articulating why a site is flagged as malicious, highlighting suspicious elements using a memory-efficient implementation of LLaMA 3.2. It utilizes a structured two-step prompt architecture to identify phishing features, generate contextual explanations, and render annotated screenshots that visually reinforce the warning.
Longitudinally implementing PhishXplain over a month on 7,091 live phishing websites, we found that it can generate warnings for 94\% of the sites, with a correctness of 96\%.
We also evaluated PhishXplain through a user study with 150 participants split into two groups: one received conventional, generic warnings, while the other interacted with PXP's explainable alerts. Participants who received the explainable warnings not only demonstrated a significantly better understanding of phishing indicators but also achieved higher accuracy in identifying phishing threats, even without any warning. Moreover, they reported greater satisfaction and trust in the warnings themselves. These improvements were especially pronounced among users with lower initial levels of cybersecurity proficiency and awareness.
To encourage the adoption of this framework, we release PhishXplain as a browser extension.

\end{abstract}




\maketitle

\section{Introduction}

Security vendors, researchers, and domain registrars have continued to improve phishing detection strategies, combining rule-based systems or heuristics~\cite{nguyen2013detecting,babagoli2019heuristic,jeeva2016intelligent,mohammad2014intelligent}, and machine learning models~\cite{sahingoz2019machine,rao2019detection,hassanpour2018phishing}to flag and remove threats before they reach users. These efforts have led to the development of robust commercial tools such as Google Safe Browsing~\cite{safebrowsing}, which demonstrate high accuracy against traditional phishing threats. 
However, when the tools fail to detect phishing websites, users are often left vulnerable, particularly those with limited technical or cybersecurity expertise~\cite{canfield2016quantifying,chen2020examination}. 
Alarmingly, research shows that even technically proficient users struggle to distinguish well-crafted phishing websites from legitimate ones~\cite{ho2024understanding}.
Browser-based anti-phishing tools usually issue standardized warnings without offering users any contextual information or explanation for why a site has been flagged~\cite{akhawe2013alice,egelman2008you,yang2015effectiveness}.
For example, Figure~\ref{fig:generic_warning} shows the generic phishing alert displayed by Google Safe Browsing (GSB), which is used by all major browsers like Chrome, Firefox, and Safari - which collectively account for roughly 95\% of global browser usage as of March 2025~\cite{globalstats:2022}.
Because these warnings lack clarity or actionable insight, users often ignore them, inadvertently putting themselves at risk. 
This challenge is further exacerbated by the increasing sophistication of phishing attacks themselves, which now leverage advanced phishing kits and generative AI tools~\cite{roy2024chatbots}, such as ChatGPT, to create evasive phishing attacks, which can remain active for hours or even days before being detected~\cite{oest2020sunrise,oest2020phishtime}. 

While limited prior work has explored the concept of contextual phishing warnings~\cite{greco2023explaining,desolda2023explanations}, existing approaches suffer from significant limitations. 
Some rely on manually crafted explanations based on visual inspection of the phishing websites~\cite{petelka2019put} - an approach that is infeasible for automatic real-time deployment. 
Others focus solely on URL features, offering minimal improvement over generic warnings~\cite{desolda2023explanations}.
In practice, transitioning these methods into dynamic, real-time systems presents several challenges.
Commercial anti-phishing tools such as GSB primarily depend on  blocklists populated by automated systems (crawlers) scanning millions of URLs or by manual user submissions from security practitioners~\cite{liang2016cracking}.
These submissions do not provide enough information that can help the tools in creating explainable warnings~\cite{oest2020phishtime}.
For example, most blocklists simply require submission of the URL. 
More recently, some providers (such as GSB) have explored fully client-side phishing detection methods that operate without blocklists. 
However, they typically use lightweight machine learning models, which tend to significantly underperform compared to blocklist-based systems~\cite{pourmohamad2024deep}.
On the other hand, more advanced research has introduced deep learning methods~\cite{lin2021phishpedia,liu2022inferring} that utilize website screenshots for phishing detection. 
However, not only are deep-learning models inherently "black-box" in nature,  their high computational demands and latency, often requiring several seconds to predict a single website makes them impractical for real-world deployment as well.
While recent advancements in post-hoc explainability techniques like LIME~\cite{ribeiro2016should} and SHAP~\cite{lundberg2017unified} have had limited success in interpreting deep learning decisions, they too are computationally intensive and produce explanations that are often too technical or abstract for the average user to understand. 

To overcome these limitations, we present PhishXplain (PXP) - a novel framework that uses a local Large Language Model (LLM) - LLaMA 3.2:3B to generate personalized, contextual phishing warnings in real-time, without imposing significant strain on the user's system resources. 
PXP augments existing browser-based anti-phishing tools by replacing generic alerts with rich, detailed explanations that highlight specific suspicious elements within a flagged website. 
LLMs have proven highly effective at generating human-readable explanations, capturing nuanced patterns, and offering contextual insights.
Recent findings also suggest that LLMs perform well in identifying phishing related features directly from a website’s source code~\cite{koide2023detecting,heiding2023devising}.
To further improve user comprehension of the warnings provided by PXP, each warning includes a screenshot of the flagged website with annotated highlights of the suspicious features, serving as visual cues. PXP operates by analyzing client-side website source code to detect user-visible phishing indicators, guided by a custom-developed lookup table and carefully tuned prompt-engineering strategies.  This enables fast, real-time generation of meaningful explanations, typically within five seconds.
In our evaluation, PXP successfully generated warnings for approximately 94\% of phishing websites encountered and accurately identified suspicious features with an accuracy of 96\%, only missing websites that relied exclusively on backend evasions (such as OAuth redirection flows or API-layer abuse) without any user-facing indicators. In these cases, PXP did not explain any suspicious elements visible on the front end. 
It is important to note that while PXP does not directly improve the underlying detection accuracy of anti-phishing tools, it plays a critical role by clearly highlighting malicious elements and explaining why they are risky, which in turn can help users internalize phishing patterns and cues.

To assess the effectiveness of PhishXplain’s explainable warnings, we conducted a controlled user study on Prolific with 150 participants, evenly split into two groups. 
One group received PhishXplain’s contextual, explainable warnings, while the other was shown generic Google Safe Browsing warnings. 
After viewing their respective warnings, participants were asked to evaluate a series of websites presented without any warnings.
We found that explainable warnings not only increased users’ immediate compliance with the warnings but also significantly enhanced their ability to detect phishing websites independently in subsequent tasks, with the improvement being especially notable among participants with lower initial technical proficiency. 
Furthermore, participants who received the explainable warnings also identified more accurate suspicious features in the website. 
Finally, they also report higher satisfaction and felt more confident in their ability to identify phishing attempts, compared to those who encountered generic GSB warnings.
The primary contributions of our work are as follows:
\begin{enumerate}
    \item We implement the first real-time, on-device approach to generating contextual phishing warnings (PhishXplain) that ensures user privacy and replaces generic phishing warnings in order to provide users with more actionable information.
    \item Using prompt-engineer and custom lookup tables, we provide an optimized implementation that demonstrates a practical, lightweight LLM deployment that has minimal resource requirements and can even run on average consumer-grade hardware.
    \item We conducted a controlled user study with 150 participants, showing that our contextual warnings significantly enhance phishing detection and user confidence, particularly among users who have low proficiency in cybersecurity.

\end{enumerate}

\section{Related Work}
Phishing is a socio-technical threat, and its success often hinges on human behavior rather than technical vulnerabilities~\cite{vayansky2018phishing}. 
Despite browsers iteratively improving SSL and site safety warnings, Akhawe and Felt~\cite{akhawe2013alice} found that users often ignore such warnings, perceiving them as either too intrusive or too vague to be meaningful, findings which were later corroborated by various other literature~\cite{junger2017priming,arachchilage2014security,abroshan2021phishing}.
This challenge is compounded by the fact that most phishing warnings offer little or no explanation of as to \emph{why} a website is dangerous. 
Alsharnouby et al.~\cite{alsharnouby2015phishing} argue that this lack of context reduces user motivation to learn and fosters over-reliance on automated tools. 
These findings align with a broader concern that users are insufficiently supported in understanding and adapting to phishing risks, especially for evasive threats, which can take hours, if not days, for anti-phishing tools to detect~\cite{oest2020phishtime}.
In enterprise settings, security awareness training is typically delivered through videos, documents, or annual workshops.
Kirlappos and Sasse~\cite{kirlappos2011security} note that such compliance-driven training is often outdated, lacks practical exercises, and is scheduled too infrequently, leading to disengagement and minimal behavior change, a finding also corroborated later by Ho et al.~\cite{ho2024understanding}. 
Puhakainen and Siponen~\cite{puhakainen2010improving} further argue that these approaches lack grounding in behavioral science and are rarely evaluated for efficacy.
These limitations have led to calls for phishing education frameworks that are not only engaging but also contextual, explainable, and reflective of real-world threats~\cite{aneke2021help}. 
Some anti-phishing models~\cite{petelka2019put,desolda2023explanations} in the recent past have tried simple rule-based explanations (for example, “This site is blocked because its URL is similar to paypal.com”), but creating a comprehensive rule set for all phishing scenarios is difficult. 
Moreover, modern phishing detectors often use machine learning classifiers that consider dozens of features (URL patterns, page content, sender reputation, etc.), which are not straightforward to communicate. 
To bridge this gap, researchers have explored explainable AI (XAI) techniques such as LIME and SHAP, which aim to identify the input features - such as words, tokens, or structural elements—that most influenced a model’s prediction~\cite{greco2023explaining,shafin2024explainable}. While these methods can offer insights into model behavior, they are typically designed for post-hoc analysis and model debugging rather than real-time, user facing explanations, and are also technically expensive.
Our work addresses these limitations by introducing a system that utilizes the natural language generation capabilities of large language models to provide dynamic, explainable warnings with feature-focused narratives that can operate in real time using limited system resources.

\section{Framework}

\begin{figure}
\centerline{\includegraphics[width=0.6\columnwidth]{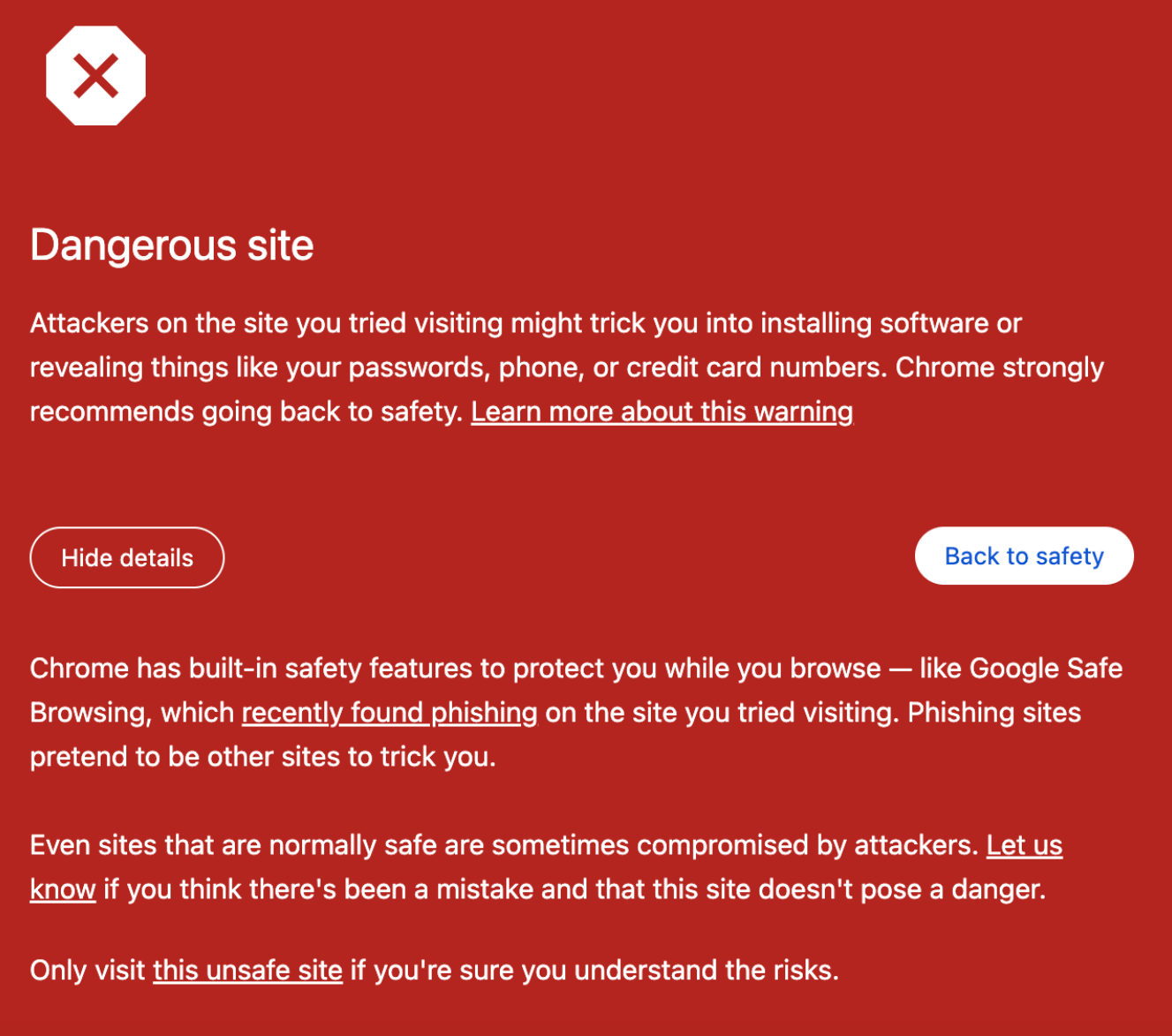}}
\caption{The generic phishing warning shown by Google Safe Browsing with no contextual information}
\label{fig:generic_warning}
\end{figure}

\begin{figure*}
\centerline{\includegraphics[width=0.8\textwidth]{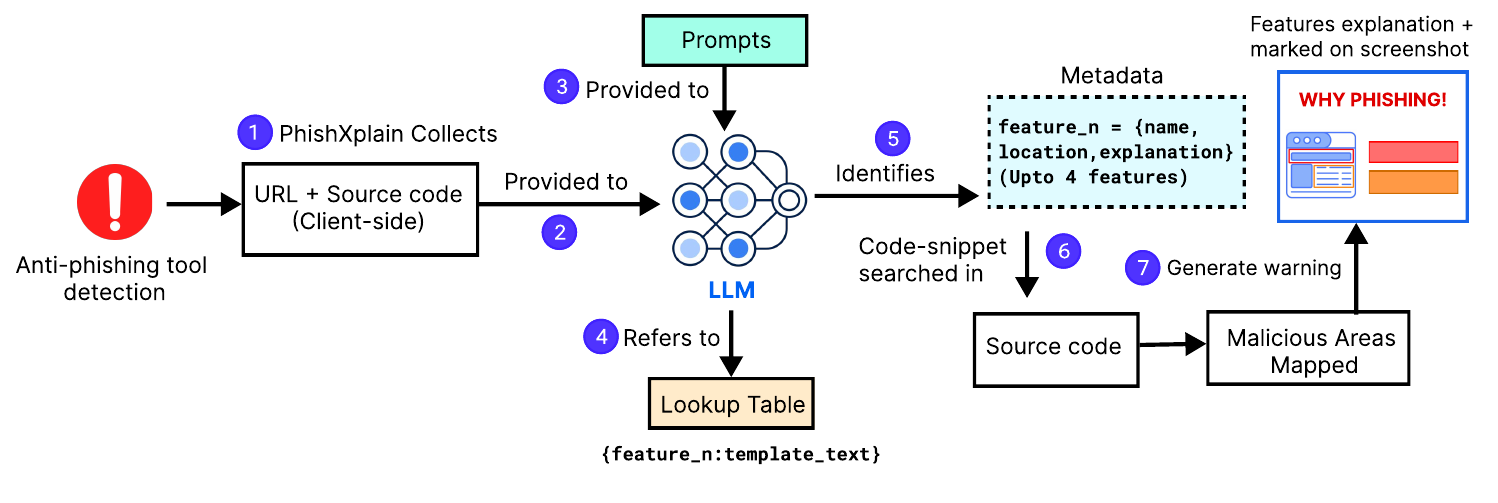}}
\caption{The PhishXplain framework}
\label{fig:framework}
\end{figure*}

\begin{figure*}
\centerline{\includegraphics[width=0.8\textwidth]{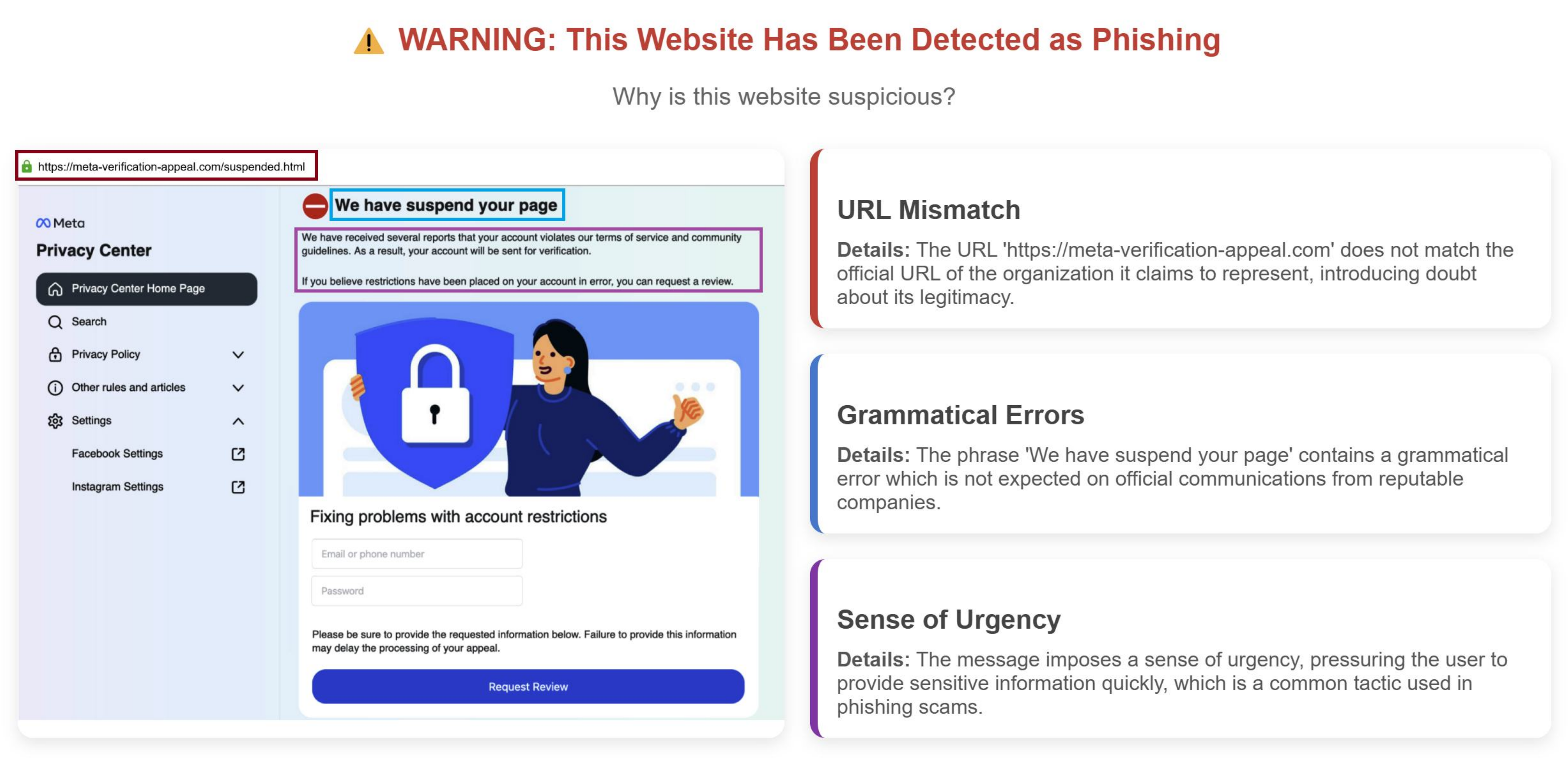}}
\caption{Example of PhishXplain's explainable blocklisting warning page showcasing contextual details about suspicious features present in the detected Facebook phishing page.}
\label{fig:example_explainable_warning}
\end{figure*}

Figure~\ref{fig:example_explainable_warning} illustrates a warning generated by PhishXplain (PXP). 
The warning contains two main components: (1) a screenshot of the flagged website and (2) a descriptive list highlighting up to four malicious features identified on the webpage. 
Each identified feature includes an explanation tailored specifically to the content of the analyzed website, outlining each malicious feature and why it is suspicious.
These features are visually color-coded and marked with bounding boxes on the website screenshot for better user comprehension.
Figure~\ref{fig:framework} provides a general overview of our framework towards generating these warnings in real-time. 
PXP functions as an additional layer of protection, activating after an initial phishing detection by the default anti-phishing tool (such as GSB).
A key motivation behind the design and optimization of PXP was to ensure that it could be seamlessly deployed on real user systems, providing meaningful phishing warnings during everyday web usage. 
In the following section, we begin by detailing the functionality each of the individual modules of our framework.

\subsection{URL and Source Code Capture} 
\label{url_and_source_code_capture}
We begin by capturing both the URL and the fully rendered client-side source code after the webpage has finished loading in the user's browser. To achieve this, we inject a lightweight script that accesses the browser’s in-memory DOM once the page is fully rendered. 
This approach ensures that we capture the version of the webpage as seen by the user, including all dynamically loaded content - such as asynchronous resources, and third-party scripts.
The capture is triggered only after the \texttt{window.onload} event fires and the network remains idle for at least 500 milliseconds, ensuring the inclusion of any late-loading components. We also log the canonicalized URL after all redirections have occurred. 
\subsection{Parsing the source-code}
\begin{figure}[H]
\centerline{\includegraphics[width=1\columnwidth]{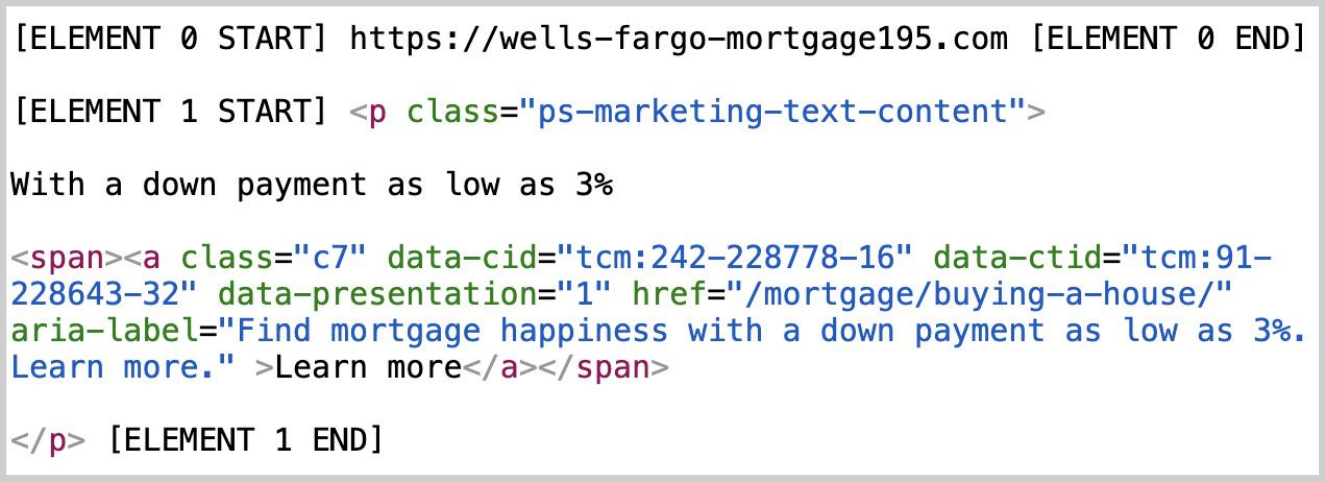}}
\caption{A \texttt{<p>} tag encapsulated by an identifier. Note the URL is appended at the start as \texttt{ELEMENT 0}. The URL is visible since it is a placeholder URL.}
\label{fig:parsing_example}
\end{figure}

To prepare the website source code for LLM-based analysis, we implement a preprocessing step that parses the HTML and programmatically wraps all user-facing elements that are commonly manipulated in phishing attacks. Specifically, we extract and isolate \emph{ten} tags categories:\texttt{<p>}, \texttt{<ol>}, \texttt{<h*>}, \texttt{<a>}, \texttt{<iframe>}, \texttt{<ul>}, \texttt{<form>}, \texttt{<button>}, \texttt{<li>}, and \texttt{<input>}. These tags were selected based on prior empirical analysis of phishing site structures, particularly the work by Roy et al.~\cite{roy2024phishlang}, which identifies these elements as frequently weaponized in phishing campaigns to deceive users or capture input.
Each of the extracted elements is then encapsulated using uniquely indexed textual delimiters of the form \texttt{[ELEMENT X START]} and \texttt{[ELEMENT X END]}, where $X \in \{1, 2, 3, \dots, n\}$ and $n$ represents the total number of identified front-end components. These delimiters are inserted directly into the source code to clearly define element boundaries. In cases where multiple identical tags are nested (e.g., a \texttt{<p>} tag within another \texttt{<p>}), the parser defaults to encapsulating only the outermost instance to preserve semantic grouping and avoid redundancy.
By marking each tag with clear delimiters, we can explicitly instruct the LLM to identify the problematic malicious tags from the source code with minimal ambiguity. Figure~\ref{fig:parsing_example} illustrates an example of how a \texttt{<p>} tag is encapsulated. On the other hand, the URL is always assigned \texttt{ELEMENT 0} and appended at the start of the source code. 

\subsection{Feature Extraction}
\label{feature_extraction}
\begin{figure}
\centerline{\includegraphics[width=0.8\columnwidth]{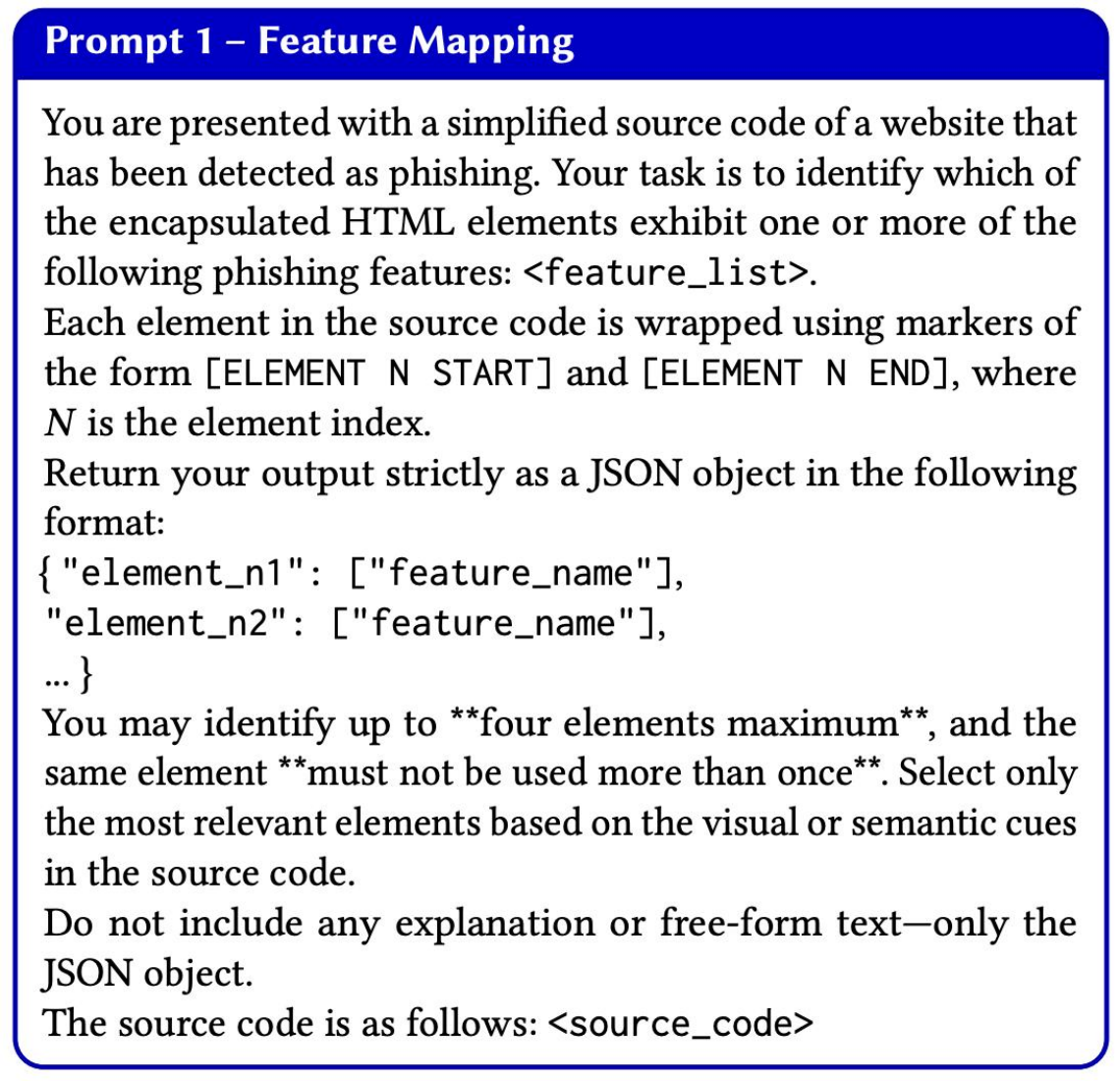}}
\caption{The first prompt to extract features from the source code}
\label{fig:prompt1}
\end{figure}

\begin{figure}
\centerline{\includegraphics[width=0.8\columnwidth]{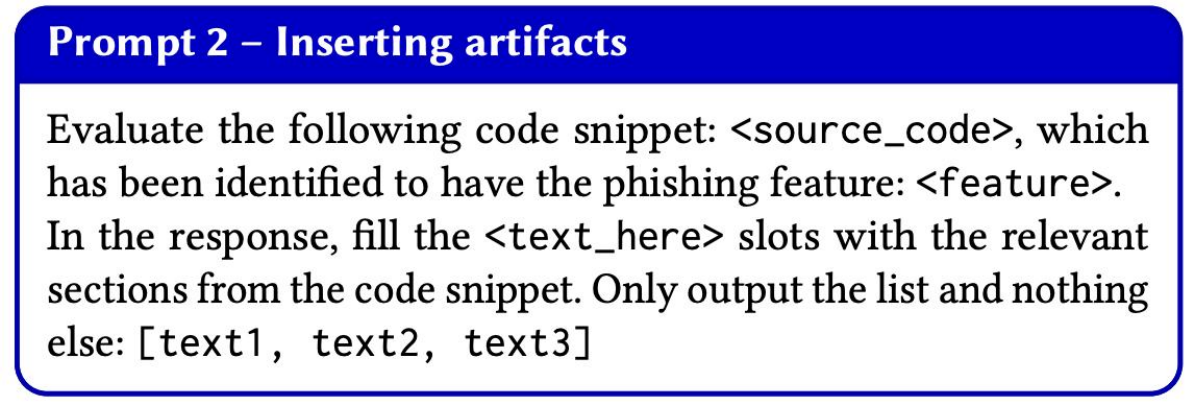}}
\caption{The second prompt to insert artifacts into the template text}
\label{fig:prompt2}
\end{figure}

The parsed source-code is then passed to the LLM through a pair of structured prompts designed to extract information about malicious features present in the website's source.
Most evasive phishing attacks are engineered to bypass anti-phishing crawlers, enabling them to stay online for extended periods. 
While this often involves sophisticated manipulations on the backend source-code, attackers are ultimately constrained by the need to \emph{display} a limited set of deceptive cues to the user to convince them to share their credentials~\cite{yuan2024adversarial}.
These include tactics such as deception, urgency, and trust exploitation by introducing spelling errors, spoofed logos, or misleading links.
However, prompting an LLM to identify these cues introduces two key challenges. 
Firstly, LLMs, especially smaller models, are known for their nondeterministic behavior and hallucinations~\cite{friel2023chainpoll,martino2023knowledge}. 
This can lead to the identification of inconsistent or even fabricated features, varying across multiple executions, even for the same website.
Secondly, such variability undermines the standardization of extracted features and makes systematic evaluation difficult.
To address these issues, we designed a two-prompt architecture that references a predefined \emph{lookup table} of phishing indicators, ensuring that the model produces standardized warning outputs.
\newline
\textbf{Lookup-table:} The Lookup table is a set of 26 common phishing features that are frequently visible to users when visiting a phishing website. 
The table consists of two columns - Feature Name and Template text.
For a particular feature, the Template text provides a sentence skeleton with multiple empty points where the LLM can insert the relevant information component specific to the phishing website it is currently evaluating. 
For example, if a feature is "Spelling errors and typos", the corresponding template text is "The website has typos such as “[fill here]” and “[fill here]”. Typos in a website which is asking you for information often indicate a phishing attempt". The model is free to fill the blank spaces with recognized artifacts. 
To develop this lookup table, we selected a random sample of 1,000 phishing websites shared on PhishTank~\cite{phishtank} in November 2023. Two coders manually evaluated these websites, snowballing newer features as they found them, coming up with the set of 26 features. Coder 1 and Coder 2 are both academic researchers, focusing on web security, phishing and LLMs. Both also designed the template text. The codebook is available here:\url{https://github.com/SayakSR/PhishXplain}.
\newline
We found that 892 websites contained three or fewer features, with 998 websites containing four or fewer features. We will publicly release the full set of coded feature labels and templates alongside our dataset, which can help researchers better understand which user-facing cues most commonly appear in phishing websites.
Based on this distribution, we configured the framework to generate a maximum of four features per warning, ensuring that the majority of phishing websites could be fully represented.
The prompts provided to the LLMs to generate the warnings, which are detailed in the next paragraph, instruct the LLM to consult with the lookup table to identify the features and the template text. 
\newline
\textbf{Prompt Design and Motivation:}
In our pipeline, we utilize two specialized prompts to ensure the LLM delivers 
controlled and context-rich explanations. 
\emph{The first prompt} focuses on identifying 
the malicious HTML tags (elements) and mapping them to a finite set of phishing features by consulting the lookup table. Then the identified malicious elements are hydrated with their respective encapsulated source code, and provide to the \emph{second prompt} which pinpoints specific suspicious artifacts from the code snippet and inserts them as artifacts in the template text. Figures~\ref{fig:prompt1} and~\ref{fig:prompt2} illustrate the prompts. It is worth noting that Prompt 2 is invoked as many times as there are malicious features. 
By limiting the model’s output to a JSON object, we avoid freeform or speculative responses~\cite{kocon2023chatgpt}, ensuring concise and predictable mappings between 
the webpage’s source code and our predefined feature set.

\subsection{Generating the warning}
\label{generating_warning}
\begin{figure}
\centerline{\includegraphics[width=0.6\columnwidth]{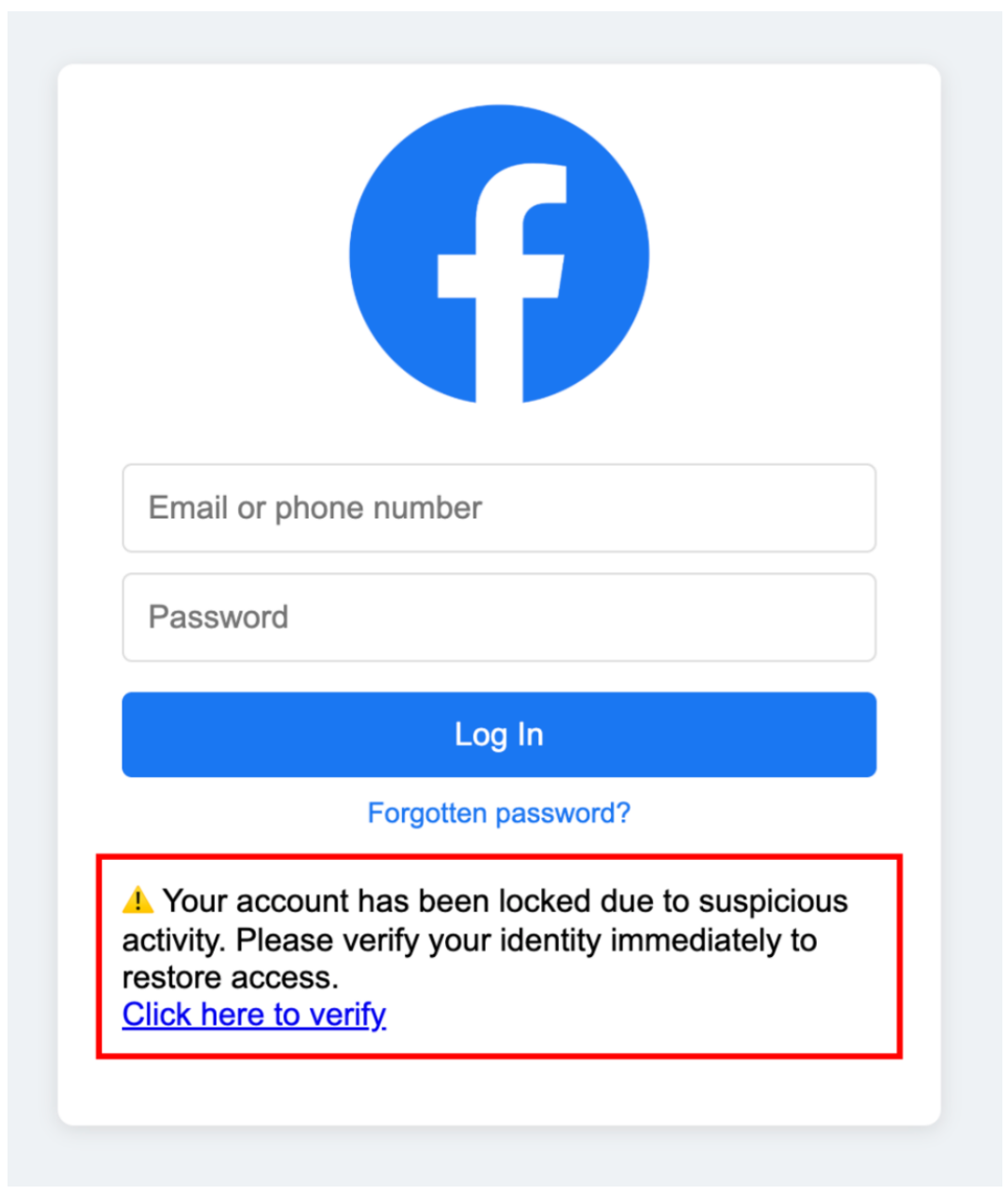}}
\caption{The original website renrendered with the malicious feature highlighted}
\label{fig:bounding_box_example}
\end{figure}

The malicious source code snippets corresponding to each identified feature (from Prompt 1) are then identified in the original source code and modified to insert a bounding box around them. 
This is implemented by injecting a CSS class that applies a colored outline to the identified code snippet. The website is then re-rendered to show the generated bounding box. Figure~\ref{fig:bounding_box_example} illustrates this in action, where the injected styling highlights the phishing content, and Figure~\ref{fig:bounding_box_example} in the Appendix shows an example where of the source code where the highlighting codeblock is inserted.

A screenshot of the annotated webpage is captured using the \textit{html2image} library and sent to the browser  along with the feature metadata, which assembles them into a user-friendly warning interface presented to the user as previously shown in Figure~\ref{fig:example_explainable_warning}. 

\subsection{Client-side application}
\label{client_side_application}
In the previous sections, we described how each component of our framework works in sequence to produce an explainable phishing warning. 
In this section, we discuss how the PhishXplain (PXP) application runs on an end-user's system. It has two modules: (1) a background service that loads the language model and manages data parsing and warning generation and (2) a lightweight Chromium extension that runs in the user's browser. 
The extension is responsible for collecting website metadata (URL and source-code) and sending it to the service, while also rendering the warning page in the browser once the response is received from the service.
\newline
\textbf{The service:} The service acts as the core engine of the PhishXplain framework. It is triggered when the browser extension detects that the built-in anti-phishing tool has flagged a webpage. At that point, the extension sends a payload containing the website's URL and the fully rendered source code to the service.
Once received, the service initiates the workflow described in Figure~\ref{fig:framework}: the parser extracts relevant information and generates prompts for the LLM, which refers to the lookup table to produce a structured feature JSON with artifacts.
This output is then used by the service to re-render the webpage, overlay visual cues such as bounding boxes, and take a screenshot for the warning interface.
To efficiently run the LLM on the user's system, the service uses Ollama~\cite{ollama}, a lightweight, containerized runtime tailored for deploying large language models locally. 
Ollama handles resource allocation, such as CPU usage and dynamically loading the model when needed, allowing for on-demand execution of the LLM without perpetually keeping it in memory.
To further optimize resource consumption, we use a 4-bit quantized variant of LLaMA 3.2:3b~\cite{meta_llama3_2_3b}, which can be loaded dynamically in under 0.28 seconds (median) when a phishing website is detected. 
Quantization significantly reduces memory usage and improves token generation speed while preserving model performance~\cite{jin2024comprehensive}. 
This design ensures that PXP remains responsive and resource-efficient, even on average consumer-grade hardware.
\newline
\textbf{The browser extension:} The extension remains dormant in the background and activates only when the browser's built-in antiphishing tool, such as GSB displays a phishing warning. 
To detect when such a warning is triggered, the extension continuously monitors the state of the current browser tab. 
It is programmed to recognize anti-phishing warnings from five widely used providers, including GSB, BitDefender TrafficLight~\cite{trafficlight}, Norton Safe Web~\cite{norton}, Avast Online Security~\cite{avastOnlineSecurity}, and Trend Micro Toolbar~\cite{avastOnlineSecurity}. 
Once a phishing warning is identified, PXP collects the URL and the fully rendered source code of the website, including client-side dependencies, and sends this payload to the background service. 
While the service processes this data to generate the feature JSON and annotated screenshot (a process that typically takes about four seconds), PXP temporarily replaces the default anti-phishing warning with a custom page that informs the user that the site has been detected as phishing and that a detailed explanation is being generated. The warnings are generated in under 5 seconds (median).
Prior research~\cite{mcalaney2020understanding} has shown that users often spend over 7 seconds reading security warnings, so this brief loading time is unlikely to disrupt the user experience.
To ensure transparency, PXP requests explicit user consent during installation to override the browser’s default phishing warnings. 
This consent prompt also includes a screenshot of the custom warning page so users know what to expect. 
For users who prefer not to overwrite their default warnings, the extension also provides an alternative: clicking the PXP icon at any time will generate the explainable warning on demand without replacing the original antiphishing alert from the browser. 
Once the service completes processing, it returns the feature JSON and annotated screenshot to the extension, which then updates the placeholder warning with the full, explainable version.


\section{Evaluating our framework}
Since our framework is designed to provide explainable phishing warnings in real time, we needed to evaluate our framework using four core aspects: \textit{Cost}, \textit{Speed}, \textit{Privacy}, and \textit{Reliability}. 
\textit{Cost} refers to both the financial expenditure and the computational resources required to run the framework. \textit{Speed} measures the latency in generating a warning for a new website. 
\textit{Privacy} checks whether any user data or website content needs to be sent to third-party servers or external APIs during inference, which can pose confidentiality risks. 
Finally, \textit{Reliability} captures how accurately and effectively the model can identify and explain suspicious elements on potentially malicious websites.
\newline
\textbf{System configuration:} To ensure our framework is accessible to a wide range of users, we evaluated its performance on a system configuration representative of an average consumer machine. 
For this, we referred to PassMark’s Q3 2024 hardware survey~\cite{pcbenchmarks}, a widely used benchmarking platform that aggregates performance metrics from thousands of users worldwide.
Based on this data, we configured a virtual machine with hardware specifications that were at or below those used by 70\% of PassMark users. 
The configuration included an Intel Core i5-11400 2.6 GHz hexa-core processor, 16 GB of RAM, and 100 GB of available SSD storage.
For testing, we used Ubuntu 22.04 LTS to streamline the process, although our framework is fully compatible with Windows 10 and later (using WSL2~\cite{microsoft_wsl_about}). 
We evaluated the framework’s performance on 100 live phishing websites sourced from PhishTank.
To minimize the risk of false positives, we selected only those entries that had been reviewed and confirmed as malicious by the PhishTank community.
\newline
\textbf{Model selection:} It is worth noting that, aside from generating the warning artifacts, the other components of the service have predictable costs and consistent speed.
For instance, capturing the URL and source code (Section~\ref{url_and_source_code_capture}) primarily depends on the website’s loading time, which is an external factor outside of the framework's control.
Meanwhile, consulting the lookup table, mapping features (Section~\ref{feature_extraction}), and re-rendering the website with bounding boxes to generate the final warning page~\ref{generating_warning} should all incur minimal overhead.
Since the generation of warning artifacts relies entirely on an LLM, selecting the right model required balancing all four key factors:\textit{ cost, speed, privacy} and \textit{reliability}.
Commercial LLMs such as ChatGPT, which are accessible via APIs, offer negligible system resource demands for the user. However, they incur a financial cost due to usage-based API pricing~\cite{openai_chatgpt_whisper_apis}, which makes large-scale deployment challenging. 
In contrast, open-source models like LLaMA 3.3 can be run locally but come with significant system resource requirements from acquiring and running high-end GPUs, and is not practical to be run on CPUs. Also from a privacy perspective, local deployment offers a clear advantage. 
Because our framework runs on the user's machine and handles URLs flagged by the browser's anti-phishing tool, it inherently processes parts of the user's browsing history. 
Ideally, this sensitive data should never leave the user's system. However, API-based commercial models transmit prompts to third-party servers, where they maybe used for model retraining purposes, thus posing significant privacy risks that local models avoid entirely.
Lastly, in terms of reliability and output quality, larger models with more parameters generally perform better~\cite{tuggener2024so,jawahar2023llm}, with more versatily in understanding and generating natural language.
However, increased parameter count also correlates with higher memory requirements. 
For example, models like LLaMA 3.3:70B require multiple GPUs to achieve practical inference speeds. 
In contrast, smaller variants trade performance for lower resource demands, though often with diminishing returns.

Thus, when selecting the LLM to be used for PXP, we aimed for the best tradeoff across all four dimensions, ensuring that the model would be cost-effective, fast enough for interactive use, respectful of user privacy, and sufficiently reliable for our application.
To do so, we primarily focus on six smaller local models: Llama 3.2:1b, Llama 3.2:3b, Mistral:7b, Gemma2:2B, Phi3:3.8B, and Phi3:14B. 
When assigned to capture complex reasoning patterns and multi-hop contextual relationships, smaller models can be particularly prone to hallucinations and drifting from task-specific objectives~\cite{huang2025survey}. 
However, prior work has also found smaller LLMs to be highly capable at focused extraction tasks, especially when guided by well-scaffolded prompts and constrained outputs.
As previously detailed in Section~\ref{feature_extraction}, our framework utilizing this strength by limiting the model’s role to a narrow, structured workflow: identifying up to four suspicious features from a fixed lookup table and populating pre-defined explanation templates using artifacts from the malicious code snippets. We will explore the effectiveness of this method later in Section~\ref{reliability}. 
For the sake of comprehensiveness, we also include two higher-end models, Gemma2:27B and Llama3.3:70B. 
However, since running these two models was not feasible on our baseline configuration due to memory limitations, we ran them on another VM with the same specifications, except for an increased system memory (96 GB RAM) to accommodate the models' requirements.
Thus, we passed each of the 100 live phishing websites through the PXP framework, running each of the eight LLMs.
We assessed each model on cost (measured as peak memory usage in MB), speed (time taken in milliseconds to generate the metadata), and performance (accuracy of the extracted features). 
We excluded the privacy metric from this evaluation, as local LLMs do not share data with any third-party servers. 
\shirin{This section is written very well. }
\subsection{Speed and memory costs}
For each model, we compute the overall \textbf{Speed} as the median \shirin{can you replace with mean?} time required by the model to generate the warning over 100 websites.
Since PXP is loaded dynamically by Ollama each time it needs to generate a warning, the speed also factors in the time required to load. 
Similarly, for memory cost, we compute the median amount of memory (RAM) utilized by the framework when generating the warning. 
The second and third columns of Table~\ref{tab:model-performance} highlights the performance of the models across the speed and memory criteria for each model. 

\subsection{Reliability Evaluation and Scoring Metric}
\label{reliability}
\begin{table*}[h]
\centering
\caption{Performance of quantized (Q4\_K\_M) models running locally via Ollama}
\resizebox{0.7\textwidth}{!}{%
\begin{tabular}{|l|c|c|c|c|c|c|c|}
\hline
\rowcolor{green!10}
\multicolumn{8}{|c|}{\textbf{Run on original configuration}} \\
\hline
\textbf{Model} & \textbf{Speed (s)} & \textbf{Memory (GB)} & \textbf{Reliability} & \textbf{CFR ($\mu \pm$ SE)} & \textbf{FMR ($\mu \pm$ SE)} & \textbf{AA ($\mu \pm$ SE)} & \textbf{CSA ($\mu \pm$ SE)} \\
\hline
LLaMA 3.2:1B  & 1.42  & 1.59  & 7.4  & 0.36 $\pm$ 0.05 & 0.07 $\pm$ 0.01 & 0.87 $\pm$ 0.03 & 0.80 $\pm$ 0.04 \\
LLaMA 3.2:3B  & 2.19  & 2.32  & 9.1  & 0.87 $\pm$ 0.06 & 0.04 $\pm$ 0.01 & 0.93 $\pm$ 0.03 & 0.88 $\pm$ 0.04 \\
Mistral:7B    & 2.84  & 5.68  & 8.8  & 0.82 $\pm$ 0.06 & 0.06 $\pm$ 0.01 & 0.91 $\pm$ 0.03 & 0.85 $\pm$ 0.05 \\
Gemma 2:2B    & 2.50  & 3.13  & 7.9  & 0.59 $\pm$ 0.05 & 0.09 $\pm$ 0.01 & 0.88 $\pm$ 0.03 & 0.78 $\pm$ 0.04 \\
Phi 3:3.8B    & 3.35  & 5.90  & 7.5  & 0.45 $\pm$ 0.05 & 0.11 $\pm$ 0.01 & 0.86 $\pm$ 0.03 & 0.80 $\pm$ 0.05 \\
Phi 3:14B     & 21.54 & 10.16 & 8.7  & 0.72 $\pm$ 0.06 & 0.05 $\pm$ 0.01 & 0.92 $\pm$ 0.03 & 0.89 $\pm$ 0.04 \\
\hline
\rowcolor{blue!10}
\multicolumn{8}{|c|}{\textbf{Configuration with more memory}} \\
\hline
Gemma 2:27B   & 33.15 & 17.13 & 9.2  & 0.87 $\pm$ 0.06 & 0.05 $\pm$ 0.01 & 0.94 $\pm$ 0.03 & 0.92 $\pm$ 0.05 \\
LLaMA 3.3:70B & 59.44 & 75.38 & 9.4  & 0.90 $\pm$ 0.06 & 0.03 $\pm$ 0.01 & 0.95 $\pm$ 0.03 & 0.94 $\pm$ 0.05 \\
\hline
\end{tabular}
}
\label{tab:model-performance}
\end{table*}

To evaluate the reliability of model-generated explanations, we conducted a manual review of the warnings produced by each model on 100 randomly selected phishing websites from our dataset. 
For every website–model pair, two expert coders were provided with the website’s URL, its rendered screenshot, and the parsed source code. Each model’s output was evaluated along the following four dimensions:

\begin{enumerate}
    \item \textbf{Correct Feature Rate (CFR):} The proportion of predicted features that were actually present on the website.
    \[
        \text{CFR} = \frac{\text{\# of correctly identified features}}{\text{total predicted features ($\leq 4$)}}
    \]
    
    \item \textbf{Feature Miss Rate (FMR):} The proportion of ground-truth features (i.e., phishing cues visible to users) that were not identified by the model.
    \[
        \text{FMR} = \frac{\text{\# of ground-truth features missed}}{\text{total ground-truth features (manually annotated)}}
    \]
    
    \item \textbf{Artifact Accuracy (AA):} Among the correctly identified features, the proportion for which the generated artifacts (e.g., brand names, typos, call-to-action text) were accurate and grounded in the HTML or visual content.
    \[
        \text{AA} = \frac{\text{\# of correctly generated artifacts}}{\text{\# of correctly identified features}}
    \]
    
    \item \textbf{Code Snippet Accuracy (CSA):} The proportion of predicted features for which the corresponding HTML snippet was correctly located. This is necessary for accurate visual highlighting on the warning screenshot.
    \[
        \text{CSA} = \frac{\text{\# of correctly matched code snippets}}{\text{total predicted features}}
    \]
\end{enumerate}

\vspace{1em}
\noindent
Using these components, we compute the final \textbf{Reliability Score} (on a scale from 0 to 10) as:\[\text{Reliability} = 10 \times \frac{\text{CFR} + (1 - \text{FMR}) + \text{AA} + \text{CSA}}{4}\]

\noindent
We provide the final score of a particular model as the median \textit{Reliability} scores obtained across all the 100 websites used in our ground-truth.
This composite score captures a model's ability to correctly detect phishing features, avoid omissions, generate grounded explanations, and accurately localize suspicious elements—factors essential for producing consistent and actionable phishing warnings. An example case study is provided in the Appendix.
\newline
\textbf{Interpretation:} Table~\ref{tab:model-performance} highlights the performance of the models across speed, memory usage, and reliability. 
Note that all models were run on the CPU as our average consumer configuration did not include a GPU.  
We found \textbf{LLaMA 3.2:3B} to be the most balanced model across all three dimensions. With a reliability score of 9.1, a modest memory footprint of 2.32GB, and a median inference time of just 2.19 seconds, it delivers consistently strong performance while remaining well within the resource constraints of the average consumer-grade system we have chosen.
In contrast, \textbf{LLaMA 3.2:1B}, although being faster (1.42 seconds) and using less memory (1.59~GB), achieved a significantly lower reliability score of 7.4, suggesting that it consistently failed to accurately identify and explain phishing cues. 
On the other hand, while other smaller models like \textbf{Mistral:7B}, \textbf{Gemma 2:2B}, and \textbf{Phi 3:3.8B} provided moderate to good reliability scores (ranging from 7.5 to 8.8), none of them outpeformed Llama3.2:3B, and they either consumed significantly more memory (such as 5.9GB for Mistral or Phi3) or took significantly longer to generate the warning. 
\textbf{Phi 3:14B} achieved a reliability score of 8.7 but required over 10~GB of memory and had a latency of 21.54 seconds, making it impractical for real-time usage. On the other hand, Gemma 2:27B and Llama3.3:70B, while having slightly higher reliability score overall (9.2 and 9.4 respectively), used significantly more memory (17.13GB and 59.44GB respectively) and needed much more time (33.15 seconds and 59.44 seconds) to generate the warning. 
Thus, based on this comparison, we chose LLaMa 3.2:3B as the final model used for deploying PhishXplain. 

\subsection{Real-world feasibility}

We also conduct a longitudinal evaluation to assess PXP's real-world performance in the real-world. It is important to note that PXP cannot generate warnings for all phishing websites detected by the anti-phishing tool, as some may lack user-facing suspicious features, such as those involving purely backend evasions.
We deployed PXP using our average consumer configuration with Google Chrome version 132.0.6834.110 over one month, from February 2\textsuperscript{nd} to March 4\textsuperscript{th}, 2025. We used Selenium's~\cite{selenium_python} automation driver to dynamically visit URLs listed and verified on PhishTank during this period. For each site, we tracked whether Google Safe Browsing (GSB) issued a detection and whether PXP could generate a corresponding explainable warning. If GSB did not flag a site at the time of first visit, we rechecked it every 10 minutes until the site became inactive.
Throughout the experiment, the browser visited 8,752 URLs, of which GSB detected 7,091. Among those, PhishXplain successfully generated explainable warnings for 6,659 websites (\(\sim94\%\)). In total, PXP produced 19,024 indicators for these warnings, with a median of 2.5 indicators per site. The system achieved a median response time of 5.2 seconds per warning and used approximately 2.7GB of memory.
To evaluate the quality of these warnings, we randomly sampled 500 of the detected websites, collectively containing 1,368 indicators, and manually assessed their accuracy. We found that 1,316 of these indicators (\(\sim96\%\)) were correctly aligned with the phishing characteristics of the websites. The average reliability score for these indicators was 9.3, which surpasses the score reported in our initial evaluation in \shirin{fix the reference:} Section~\ref{reliability} (Table~\ref{tab:model-performance}). 
\section{User Study}
Through the previous sections of this paper, we introduced, developed, and tested PhishXplain as a lightweight, scalable tool designed to provide contextual phishing warnings to end users. However, it remains crucial to determine whether PXP's warnings can improve users’ ability to spot phishing websites. Without empirical evidence demonstrating that users attend to, understand, and ultimately benefit from contextualized explanations, any theoretical advantages of PXP would not be beneficial in the real world. 
To that end, this section presents a controlled between subjects user study comparing PhishXplain’s explainable warnings to traditional, generic warnings. Participants evaluated a mix of phishing and benign websites, each accompanied by either a generic or explainable warning. By analyzing their accuracy, confidence, and perceived helpfulness of the warnings, we investigate whether PXP’s explainability leads to measurable improvements in phishing detection behavior.  
Our study consists of two groups: Group A, where participants see PXP's explainable warnings for four phishing websites, and then evaluate the screenshots of eight websites (four phishing and four benign); and Group B, where participants see generic warnings (modeled after the standard warning from GSB, see~\ref{fig:generic_warning}) for the same four phishing websites as Groyp A, and also evaluate the same eight websites. 
First, participants complete a background questionnaire designed to measure their cybersecurity proficiency. 
Based on their performance, participants are categorized into low, medium, or high proficiency levels. This stratification enables us to explore how the effectiveness of explainable warnings may vary across different levels of user expertise. In the following section, we introduce the research questions that we answer through this study and their corresponding hypotheses. Table~\ref{tab:study_data_summary} also shows the data types collected for each survey section.
\subsection{Research Questions and Hypotheses}
Through this user study, we answer the following four research questions and evaluate them based on corresponding hypotheses:
\noindent \textbf{RQ1: Do explainable security warnings help users more accurately identify phishing websites compared to generic phishing warnings?} \\
We evaluate whether participants who receive explainable warnings are better at correctly identifying both phishing and benign websites.

\tcbset{
  colback=gray!10,
  colframe=gray!30,
  arc=1mm,
  boxrule=0.3pt,
  left=1mm,
  right=1mm,
  top=1mm,
  bottom=1mm,
  enhanced,
  sharp corners
}

\begin{tcolorbox}
\noindent \textit{\textbf{H1:} Participants who view the explainable warning (Group A) will exhibit higher overall accuracy in classifying websites than those who view the generic warning (Group B).}
\end{tcolorbox}

\vspace{1em}
\noindent \textbf{RQ2: How do different warning types influence users’ confidence and satisfaction with the warnings?} \\
We evaluate if explainable warnings make participants feel more confident about their decisions and whether they find these warnings more helpful overall.

\begin{tcolorbox}
\noindent \textit{\textbf{H2:} Participants who receive explainable warnings will report higher confidence in their decisions than those receiving generic warnings.}

\noindent \textit{\textbf{H3:} Participants in the explainable warning group will rate the warnings as being more helpful than those in the generic warning group.}
\end{tcolorbox}

\vspace{1em}
\noindent \textbf{RQ3: Does user proficiency in cybersecurity affect how helpful or effective the warning type is?} \\
We examine whether lower-proficiency users benefit more from explainable warnings compared to higher-proficiency users, both in terms of performance and perceived helpfulness.

\begin{tcolorbox}
\noindent \textit{\textbf{H4:} The benefit of the explainable warning will be more pronounced for lower-proficiency participants than for higher-proficiency participants.}

\noindent \textit{\textbf{H5:} Lower-proficiency participants will rate the explainable warning as more helpful than higher-proficiency participants.}
\end{tcolorbox}

\vspace{1em}
\noindent \textbf{RQ4: Do explainable warnings improve users’ ability to recognize and articulate phishing cues?} \\
We check whether participants in the explainable warning group are better at identifying the specific features that make a website look suspicious when asked to explain their reasoning.

\begin{tcolorbox}
\noindent \textit{\textbf{H6:} Participants in the explainable warning group will more frequently identify correct phishing cues than those in the generic warning group.}
\end{tcolorbox}

\subsection{Study Components}
\textbf{Sample size estimation and recruitment:} 
To determine the appropriate sample size for our user study, we conducted a power analysis using standard sample size estimation methods for a two-group between-subject design. Assuming a medium effect size (\( d = 0.50 \)), a significance level of \( \alpha = 0.05 \), and a desired power of 0.80, we used the formula \( N_{\text{total}} = 2 \times \left( \frac{Z_{1-\alpha/2} + Z_{1-\beta}}{d} \right)^2 \). Substituting the standard Z-scores for a two-tailed test (\( Z_{1-\alpha/2} = 1.96 \), \( Z_{1-\beta} = 0.84 \)), the required sample size comes out to \( N_{\text{total}} = 2 \times (5.6)^2 = 2 \times 31.36 = 62.72 \), or approximately 63 participants. 
Thus, we recruited 150 participants via Prolific, an online crowdsourcing platform widely used for academic research, applying the following eligibility criteria: participants had to be based in the United States, at least 18 years of age, and possess a minimum approval rate of 95\% on prior studies. The study lasted approximately 25 minutes, and participants received \$5 upon successful completion, which aligns with Prolific’s guidelines for fair hourly compensation~\cite{prolific_informed_consent_2023}. To ensure data quality, we embedded two attention-checking questions and excluded responses from two participants who failed to answer them correctly during the study. We extracted the demographic data for the participants from Prolific itself. 
\newline

\textbf{Cybersecurity Proficiency}
In RQ3, we examine how participants' cybersecurity proficiency influences the effectiveness of explainable phishing warnings.
We do not ask users to self-report their technical proficiency as prior work has consistently shown that such self-perceptions can be unreliable, with individuals frequently overestimating their abilities~\cite{mahmood2016people,ament2017ubiquitous}.
Instead, we adopt the Human Aspects of Information Security Questionnaire (HAIS-Q) scaled developed by Parsons et al~\cite{parsons2017human}, an established questionnaire used to assess cybersecurity awareness and secure behavior across three key domains - knowledge, attitudes, and self-reported practices. For each question, participants indicate their agreement on a 5-point Likert scale from "Strongly Disagree" to "Strongly Agree" (e.g., whether it is risky to open attachments from unknown senders). Specifically, we adopted 11 items from the HAIS-Q questionnaire, including six from the Email Use subscale and five from the Internet Use subscale, taking inspiration from the study by Schoni et al.~\cite{schoni2024you}, who also determined cybersecurity proficiency in the context of phishing threats. 
Based on the correct behaviour for each question, responses were scored from 0 (for the worst answer) to 2 (for the best answer) over 0.5-point increments. Thus, the maximum possible in this section was 22 points.
Participants also answered nine multiple-choice questions designed to evaluate practical phishing recognition skills, such as identifying risky attachments and suspicious messages. Schoni et al.~\cite{schoni2024you} also influenced these questions. For each question, the correct answer was awarded 2 points, thus totaling a maximum phishing knowledge score of 18.
Thus, the maximum combined proficiency score across the 20 questions (11 from HAIS-Q and 9 for partical phishing knowledge) was 40. Based on the total scores received by the participants, we assigned each of them into three proficiency tiers inspired by Schoni et al.'s work~\cite{schoni2024you}: Low Proficiency: score < 29, High Proficiency: score > 36 and Intermediate Proficiency: all others.
\newline
\textbf{Phishing warnings:} After completing the cybersecurity proficiency questionnaire, participants were shown screenshots of four phishing websites, each accompanied by a corresponding warning. This section marks the primary divergence in content between Group A and Group B. Participants in Group A were presented with explainable warnings (similar to Figure~\ref{fig:example_explainable_warning}) that highlighted specific malicious features detected on the website and also annotated these features directly on the website screenshot (embedded in the warning). In contrast, participants in Group B were shown a generic Google Safe Browsing (GSB) warning \shirin{fix the reference:} (Figure~\ref{fig:generic_warning}), which simply flagged the website as potentially deceptive without offering any reasoning.
The four phishing websites used for the warnings were taken from PhishTank, imitating popular organizations such as Meta, Wells Fargo and Netflix, exhibiting common deceptive patterns found in phishing websites such as excessive/inappropriate information requests, unrealistic claims, IDN homographs, third-party hosting, etc.
These websites are deliberately chosen such that, in Group A, the explainable warnings highlight specific phishing indicators which then appear, in various forms, within the set of four phishing websites that participants later evaluate without any warnings (See next section). These cues would not be present for Group B participants who receive a generic warning. This design allows us to identify whether they can generalize the knowledge shown in the explainable warnings to new, unseen phishing websites.
For both groups, after viewing each warning, participants were asked why they thought the website was phishing. They were allowed to revisit the warning before submitting their answer. This encouraged participants to actively engage with the warning content rather than passively accept it, which can promote learning of the features (for Group A participants).
\newline
\textbf{Website Assessments:} After viewing the phishing warnings, participants were asked to evaluate a set of eight websites, four phishing and four benign. These websites were presented without any accompanying warnings, requiring participants to rely solely on their judgment. The phishing websites were carefully selected from PhishTank and were chosen to contain malicious features similar to those highlighted in the warning phase (for Group A). This design reinforces the learning objective of the explainable warnings, enabling us to assess whether participants could transfer their understanding of phishing cues to new, unassisted evaluations.
\shirin{fix the reference:} 
\begin{table}[ht]
\centering
\caption{Malicious websites and the phishing features they exhibit. }
\label{survey_website_info_short}
\resizebox{0.99\columnwidth}{!}{%
\begin{tabular}{p{3.5cm}|p{8.5cm}}
\toprule
\textbf{Website (Target)} & \textbf{Phishing Features} \\
\midrule
Warning-1 (Meta) & Suspicious URL, Grammatical errors, Sense of urgency \\
Warning-2 (Wells Fargo) & Requests sensitive/inappropriate information, Hosted on 3rd party domain \\
Warning-3 (YouTube) & IDN homograph (Cyrillic URL), Requests sensitive/inappropriate information, Unrealistic claim \\
Warning-4 (Adobe) & Poor design, Unusual login request \\
Phishing-1 (Netflix) & IDN homograph (Cyrillic URL), Hosted on 3rd party domain \\
Phishing-2 (Instagram) & Suspicious URL, Sense of urgency \\
Phishing-3 (eBay) & Suspicious URL, Unrealistic claim, Requests sensitive/inappropriate information \\
Phishing-4 (Bank of America) & Suspicious URL, Hosted on 3rd party domain, Requests sensitive/inappropriate information \\
\bottomrule
\end{tabular}%
}
\end{table}
Table~\ref{survey_website_info_short} provides a brief summary of the phishing indicators embedded in each of the websites. 
Unlike some earlier phishing user studies that used obviously fake or poorly constructed designs, we intentionally excluded such examples. Modern phishing attacks increasingly rely on professionally crafted phishing kits, resulting in more realistic-looking websites. Including poorly designed phishing sites would not reflect the current threat landscape as prior work by~\cite{yuan2024adversarial} et al. has shown that users perform well at identifying such low-effort phishing attempts.
After viewing each website screenshot, participants were asked to (1) indicate whether they believed the website was phishing or benign, (2) briefly explain the reasoning behind their decision, and (3) rate their confidence on a 5-point Likert scale ranging from "Not at all confident" to "Extremely confident". This design allows us to not only quantify detection accuracy but also gain insights into participants' decision-making processes, i.e., what features they recognized in the website, and also assess the confidence levels associated with their judgments. 
\newline
\textbf{Participant Satisfaction:}
After the end of the study, the participants were asked to rate how helpful the warnings were in guiding their phishing detection decisions, using a 5-point Likert scale ranging from “Not at all helpful” to “Very helpful.” They were also asked how their confidence in identifying phishing websites had changed after viewing the warnings, with response options ranging from “Much worse than before” to “Much better than before.” This was to assess how participants perceived the usefulness of the security warnings.

\begin{table}[ht]
\centering
\caption{Overview of data types collected in each section of the user study}
\renewcommand{\arraystretch}{1.2}
\resizebox{0.9\columnwidth}{!}{
\begin{tabular}{p{5cm}|p{6.3cm}}
\hline
\textbf{\rule{0pt}{1.3em}Section (No. of Questions)} & \textbf{\rule{0pt}{1.3em}Data Collected} \\
\hline
\rule{0pt}{1.3em}Demographics (4) & Categorical variables \\
\hline
\rule{0pt}{1.3em}Cybersecurity Proficiency (20) & 
Likert-scale responses (HAIS-Q) \newline
Multiple-choice answers (phishing knowledge) \newline
Numeric proficiency score (0–40) \\
\hline
\rule{0pt}{1.3em}Phishing Warnings (4) & Open-ended text responses \\
\hline
\rule{0pt}{1.3em}Website Assessments (8) & 
Binary phishing/benign label \newline
Open-ended justification \newline
Confidence rating (5-point Likert scale) \\
\hline
\rule{0pt}{1.3em}Participant Satisfaction (2) & 
Helpfulness rating (5-point Likert scale) \newline
Confidence change (5-point Likert scale) \\
\hline
\end{tabular}
}
\vspace{0.5em}
\label{tab:study_data_summary}
\end{table}

\subsection{Analysis} In this section, we detail the analysis methods used for each hypothesis and the subsequent results, organized by research question. For each hypothesis, we begin with descriptive statistics to summarize the key trends in the data, which is followed by inferential statistical analysis to test for significance. 
\newline
\textbf{Normality and choice of tests:} Before selecting appropriate statistical tests for our hypotheses, we assessed the distributional properties of the key variables used in our analysis: detection accuracy (H1), average confidence scores (H2), perceived helpfulness ratings (H3 and H6), and number of phishing features correctly identified (H4). For each variable, we conducted a Shapiro–Wilk test to evaluate normality within both participant groups (Group A and Group B). Across all variables and groups, the Shapiro–Wilk tests returned $p$-values below 0.05, indicating significant deviation from normality. 
Thus, we adopted the non-parametric Mann–Whitney U tests) for our between-group comparisons wherever appropriate. On the other hand, for factorial hypotheses involving multiple categorical factors (e.g., H4 and H5), we used two-way ANOVA models due to their ability to test interaction effects between warning type and proficiency. 
\newline 
\textbf{Bonferroni correction:} To control for potential Type~I errors, given the number of hypotheses tested, we applied Bonferroni corrections across families of related outcomes. For performance-related hypotheses (H1, H2, H3, H4), we used a corrected significance threshold of $\alpha = 0.0125$. For perception-related hypotheses (H5, H6), the threshold was set at $\alpha = 0.025$. This family-wise correction ensures that observed effects are statistically robust and not due to chance.

\textbf{RQ1: Do explainable warnings improve users’ accuracy and efficiency in detecting phishing websites?}

\begin{table}[ht]
\centering
\resizebox{0.7\columnwidth}{!}{%
\begin{tabular}{lcc}
\toprule
\textbf{Website} & \textbf{Group A (n=75)} & \textbf{Group B (n=75)} \\
\midrule
Phishing-1  & 68 (90.67\%) & 39 (52.00\%) \\
Phishing-2  & 67 (89.33\%) & 49 (65.33\%) \\
Phishing-3  & 59 (78.67\%) & 41 (54.67\%) \\
Phishing-4  & 67 (89.33\%) & 52 (69.33\%) \\
Benign-1    & 64 (85.33\%) & 53 (70.67\%) \\
Benign-2    & 67 (89.33\%) & 62 (82.67\%) \\
Benign-3    & 33 (44.00\%) & 30 (40.00\%) \\
Benign-4    & 58 (77.33\%) & 51 (68.00\%) \\
\bottomrule
\end{tabular}}
\caption{Absolute and percentage of participants who correctly evaluated each of the unguided websites}
\label{tab:website_accuracy}
\end{table}

For \textbf{H1}, considering the eight websites where participants were unguided (i.e., they did not see an explainable warning), we find that participants in Group A, on average, correctly identified 6.44 websites (median=7). In contrast, participants in Group B, on average, correctly identified 5.16 websites (median=5). 
Table~\ref{tab:website_accuracy} illustrates the participants' accuracy for Group A and Group B for each of the eight websites for which they did not receive warnings. Across the phishing websites, Group A consistently outperformed Group B, with notably higher correct accuracy rates on all four phishing sites. For example, 90.67\% of Group A participants correctly identified Phishing-1, compared to only 52\% for Phishing-3. Group A had an accuracy of 78.67\%, while Group B had only 54.67\%. 
However, for the benign websites, the accuracy rate was closer. While Group A had slightly higher accuracy on Benign-1, Benign-3, and Benign-4, Group B slightly outperformed Group A on Benign-2. This suggests that while Group A participants showed a clear advantage in detecting phishing attempts, both groups performed similarly when recognizing legitimate websites.
Now, to statistically evaluate whether explainable warnings enhanced phishing detection accuracy, we calculated the number of correctly classified websites (out of eight) for each of the Group A and Group B participants and ran a Mann-Whitney U test to test for statistical significance. Overall, participants in Group A were significantly more likely to identify websites correctly than participants in Group B ($p < 0.01$). When only considering the four phishing websites, this also remains true ($p < 0.01$); however, when considering the benign websites, the difference between Group A and Group B was not statistically significant ($p = 0.055$). This result is marginal and suggests a possible trend but is not strong enough to confidently conclude that explainable warnings were more beneficial to Group A participants correctly identifying benign websites than Group B participants who received generic warnings. 

\begin{tcolorbox}[colback=gray!10, colframe=black, boxrule=0.5pt]
Explainable warnings significantly improve the detection accuracy for phishing websites but do not necessarily prevent the user from misidentifying benign websites compared to generic warnings.
\end{tcolorbox}

\textbf{RQ2: How do different warning types influence users’ confidence and satisfaction with the warnings?}

For \textbf{H2}, to assess whether explainable warnings increased the users’ confidence, we computed the average confidence score (on a 5-point Likert scale) across all evaluations per participant and again compared the values using the Mann-Whitney U test. Participants in the explainable group reported significantly higher confidence (average = 3.83) than those in the generic group (average = 2.77), with $p<0.015$. For \textbf{H3}, to evaluate perceived helpfulness, we compared participants’ post-survey ratings using the Mann-Whitney U test. For Group A, 49 participants found the explainable warning ``Very helpful," whereas 24 participants found it ``Somewhat helpful." Only 1 participant found the warnings ``not very helpful." On the other hand, only 7 participants found the warning ``Very helpful", with 25 finding them ``somewhat helpful,'' whereas 11 participants were ``Neutral." However, the remaining 32 participants found it ``not very helpful." Neither Group A or Group B participants answered the lowest score, i.e., ``Not very helpful." 
Explainable warnings were rated as significantly more helpful (average = 4.64) than generic warnings (average = 3.09), with $p < 0.01$. This indicates that users found detailed warnings to be more transparent and trustworthy. 

\begin{tcolorbox}[colback=gray!10, colframe=black, boxrule=0.5pt]
Participants reported higher confidence and perceived helpfulness when warnings explained why a site was suspicious, enhancing both trust and engagement.
\end{tcolorbox}

\textbf{RQ3: Does user proficiency affect how helpful or effective the warning type is?}
Based on our scoring threshold for the 20 HAIS-Q and phishing awareness questions, we found 29 participants (16 from A, 13 from B) who were low proficient, 84 participants who were intermediate proficient (45 from A, and 49 from B), and 26 who were High proficient (14 from A, and 12 from B). 
For \textbf{H4}, we used a two-way factorial ANOVA to evaluate whether user proficiency moderated the effectiveness of explainable warnings. There were significant main effects of both warning type and proficiency, and a significant interaction ($F(2,147)=6.32$, $p<0.005$). Explainable warnings yielded the most significant gains for low-proficiency participants, whose accuracy rose from a mean of 5.0 (generic group) to 6.8. In contrast, high-proficiency participants performed similarly across groups (7.3 vs. 7.5).
For \textbf{H5}, to determine whether proficiency influenced perceived helpfulness, we conducted another two-way ANOVA, finding a significant interaction between warning type and proficiency group ($F(2,147) = 4.10$, $p = 0.015$). Lower proficiency rated explainable warnings as significantly more helpful than high-proficiency users.
\newline

\begin{tcolorbox}[colback=gray!10, colframe=black, boxrule=0.5pt]
Explainable warnings disproportionately benefited low-proficiency users, improving both performance and perceived helpfulness and helping bridge the cybersecurity expertise gap.
\end{tcolorbox}
\textbf{RQ4: Do explainable warnings improve users’ ability to recognize and articulate phishing cues?}

To evaluate the features (cues) recognized by the participants when viewing the warning and also when evaluating the website, our two coders analyzed participants’ open-ended justifications using features from our lookup table (Section~\ref{feature_extraction}) and assigned features to each response that were correct. The coders had a Cohen's Kappa inter-rater agreement of 0.72, indicating substantial agreement, and all disagreements were resolved. We first summed the distinct number of features across the four initial warning questions, i.e., where the participants were asked to identify why a warning was being shown for the phishing website. We found that respondents in Group A, on average, pointed out 9.19 features (median=9), while those in Group B pointed out 6.24 features (median=6). 
For \textbf{H6}, we compare the totals across the participants using a Mann-Whitney U Test, we found that participants in Group A were significantly more likely to identify a correct feature ( $p<0.01$) after seeing the explainable warning, compared to Group B participants who saw the generic warning which did not provide them with any details regarding why the website was phishing. 
Then, we focus on the eight websites the participants evaluated without any warnings. We first identify the correct feature rate for the four phishing websites in this set and find that Group A participants, on average, identified 7.84 features \shirin{report means and std errors}(median=8). In contrast, Group B participants identified 5.45 features (Median=5). Performing a Mann-Whitney U test, we again find that respondents in Group A were significantly more likely to identify correct features in the unguided phishing websites than participants in Group B ( $p < 0.01$). 
On the other hand, we also check if participants had stated any malicious features for the four benign websites, considering them as them pointing out incorrect features. In this case, we find that Group A participants, on average, marked 2.41 \shirin{report means and std errors} (median=2) features incorrectly for benign websites, while Group B participants marked on average 3.23 (median=3) features incorrectly. Performing the Mann-Whitney U test, we \textit{do not find statistical significance} ($p = 0.129$), suggesting that there may be a trend toward Group B participants being more likely to mark benign elements as suspicious, but this pattern is not strong enough to draw firm conclusions from.
This finding aligns with our findings in RQ1, where we see that while explainable warnings improve user accuracy in identifying phishing websites, they do not necessarily prevent users from marking benign websites as phishing.

\begin{tcolorbox}[colback=gray!10, colframe=black, boxrule=0.5pt]
Explainable warnings improved participants’ recognition of malicious features for phishing websites but did not necessarily prevent them from misidentifying such features in benign websites.
\end{tcolorbox}

\subsection{Interpretation}
Our controlled user study comparing PhishXplain with generic phishing warnings provides strong evidence that contextualized warnings significantly improve phishing detection by end-users, especially among those with lower cybersecurity proficiency. Participants who received PhishXplain’s detailed warnings not only identified more phishing websites correctly but also reported greater confidence in their decisions and rated the warnings as more helpful overall. By contrast, participants who received generic warnings often struggled to determine exactly what was suspicious about a site, leading to lower accuracy and smaller gains in confidence.
From a proficiency standpoint, explainable warnings proved especially beneficial for lower-proficiency participants, lifting their phishing detection performance to nearly the same level as higher-proficiency users. This can be because less experienced users often lack the well-formed mental models that security-savvy individuals rely on~\cite{zielinska2015exploring} to spot hidden cues- for instance, subtle URL manipulations, unusual information requests, or grammatical errors. By visually highlighting these risky elements and providing contextualized rationales, PhishXplain effectively ``fills in” these missing mental models in real time. As a result, novices learn why a site might be malicious, and they carry that insight over to unguided websites where no warning is present.
Despite these advantages, our results show that explainable warnings did not reduce the rate at which participants mislabeled benign websites. However, there was also no evidence that the warnings increased false positive detection. We assume that since PhishXplain focuses on showcasing potentially malicious cues for phishing websites without simultaneously reinforcing signs of legitimacy for benign ones, users may remain uncertain when no obvious phishing red flags are present, tending to err on the side of caution rather than receive explicit reassurance that a site is safe. This phenomenon of risk-averse user behavior aligns with prior research on security prompts~\cite{sharma2021impact,malkin2017personalized}, which similarly reports that users often opt for defensive responses when uncertain.

\begin{table}[ht]
\centering
\renewcommand{\arraystretch}{1.2}
\caption{Summary of the participant-level data points used for each hypothesis.}
\resizebox{0.98\columnwidth}{!}{
\begin{tabular}{p{0.8cm}|p{4.7cm}|p{5.5cm}}
\hline
\textbf{Hyp.} & \textbf{What is Tested} & \textbf{Data Point (Per Participant)} \\
\hline
H1 & Detection accuracy across websites & Total correct classifications (0–8) \\
\hline
H2 & Confidence in decisions & Avg. confidence rating across 8 websites (Likert 1–5) \\
\hline
H3 & Accuracy × Proficiency interaction & Accuracy score (0–8), grouped by proficiency (Low/Med/High) \\
\hline
H4 & Recognition of phishing cues & Total number of distinct features mentioned (coded from open-ended responses) \\
\hline
H5 & Evaluation time & Avg. time taken across 8 websites (in seconds) \\
\hline
H6 & Perceived warning helpfulness & Post-survey helpfulness rating (Likert 1–5) \\
\hline
H7 & Helpfulness × Proficiency interaction & Helpfulness rating (Likert 1–5), grouped by proficiency \\
\hline
\end{tabular}
}
\label{tab:hypothesis_data_points}
\end{table}



\section{Limitations and Future work}
Our user study was conducted in a controlled setting and evaluated only the immediate impact of contextual warnings on user behavior. 
While this design allowed for an initial assessment of PhishXplain’s advantages over generic alerts, it does not fully capture how users might respond to repeated exposure or real-life distractions in everyday browsing. 
We plan to address this by conducting a longer-term field study where participants install and use PXP continuously in their day-to-day browsing for several weeks or months, which will allow us to observe how sustained use and repeated exposure to explainable warnings influence user learning, trust, and security habits over time.
Second, although PXP’s detailed cues significantly improved phishing detection rates, they did not enhance participants’ accuracy when judging benign websites. 
This imbalance suggests a risk that emphasizing malicious cues alone could heighten suspicion to the point where legitimate sites are sometimes misjudged. 
As a next step, we plan to incorporate optional positive indicators for benign sites, rather than framing them as “warnings” per se. 
These cues could appear in a pop-up or extension icon,  which users can click to see the benign signals.


\section{Conclusion}
In this paper, we introduced PhishXplain (PXP), a lightweight and scalable framework that complements anti-phishing tools by replacing generic phishing warnings with contextual, feature-rich explanations regarding why a website is phishing. 
The framework uses a locally hosted large language model (Llama 3.2:3B) to identify malicious features within a website and  highlight them to the user through textual and visual means.
Through careful prompt engineering, feature lookups, and on-device deployment, PXP can generate and display the warnings in real-time inference very quickly. It also uses minimal system resources and is privacy-focused, making it suitable for practical, everyday use, even on average consumer-grade hardware. 

To test PXP's efficiency, we performed a controlled user study. Participants exposed to PXP’s explainable warnings were substantially more accurate at identifying new phishing pages and reported higher confidence in their decisions compared to those who were exposed to generic Google Safe Browsing style warnings. 
This effect was especially prominent in users with lower cybersecurity proficiency, suggesting that PXP can positively address the knowledge gap required to identify phishing attacks in the wild. 
Participants who viewed PXP's warnings were also more adept at identifying correct phishing cues in websites and showed greater satisfaction with the overall warning experience. 
Collectively, these results underscore that explaining why a site is suspicious can boost user comprehension and preparedness against phishing threats online, thus sustaining the necessity of PhishXplain. To help researchers and users alike, we release PhishXplain as a browser extension at \url{https://github.com/SayakSR/PhishXplain}, licensed for academic and non-commercial use.

\bibliographystyle{ACM-Reference-Format}
\bibliography{refs}

\section*{Survey Questionnaire}

\subsection*{Cybersecurity Habits}

\begin{enumerate}

    \item Please indicate your level of agreement with the following statements (Strongly Disagree to Strongly Agree):
    \begin{itemize}
        \item It’s risky to open an email attachment from an unknown sender.
        \item It’s always safe to click on links in emails from people I know.
        \item Nothing bad can happen if I click on a link in an email from an unknown sender.
        \item I don’t open email attachments if the sender is unknown to me.
        \item If an email from an unknown sender looks interesting, I click on a link within it.
        \item I don’t always click on links in emails just because they come from someone I know.
        \item It can be risky to download files on my work computer.
        \item Just because I can access a website at work doesn’t mean it’s safe.
        \item If it helps me do my job, it doesn’t matter what information I put on a website.
        \item I download any files onto my work computer that help me get my job done.
        \item When accessing the Internet at work, I visit any website I want.
    \end{itemize}
\end{enumerate}

\subsection*{Phishing Knowledge Assessment}

\begin{enumerate}
    \item What is phishing?
    \begin{itemize}
        \item A. A type of malware that infects computers
        \item B. A deceptive advertising tactic
        \item C. A social engineering trick to buy products
        \item D. An online scam to steal sensitive information (Correct)
    \end{itemize}

    \item What are the risks of phishing?
    \begin{itemize}
        \item A. Identity theft
        \item B. Financial loss
        \item C. Loss of personal data
        \item D. Malware infection
        \item E. All of the above (Correct)
    \end{itemize}

    \item Which sentence would most likely appear in a phishing email?
    \begin{itemize}
        \item “As a valued customer, we’re giving you a special discount! -90\% on all our offers, click here to view more!” (Correct)
    \end{itemize}

    \item Which file looks legitimate?
    \begin{itemize}
        \item A. paypal\_account\_details.exe
        \item B. vacation\_photos.pdf.exe
        \item C. todays\_notes.txt (Correct)
        \item D. bank\_invoice.scr
    \end{itemize}

    \item What should you do if you suspect phishing?
    \begin{itemize}
        \item A. Ignore it
        \item B. Ask co-workers for opinions
        \item C. Report it so the organization can investigate (Correct)
        \item D. Open it to check if it looks legitimate
    \end{itemize}

    \item What is common content in phishing emails?
    \begin{itemize}
        \item A. Security alert of suspicious login
        \item B. Ads for weight loss supplements
        \item C. Threats of account deactivation or legal action if immediate action is not taken (Correct)
        \item D. None of the above
    \end{itemize}

    \item Why is googleaccountsupportgsupport.com suspicious?
    \begin{itemize}
        \item A. The domain address should contain google.com (Correct)
        \item B. It should contain “no-reply”
        \item C. It should be a gmail.com address
        \item D. Capital letters are required in company names
    \end{itemize}

    \item Which of the following messages is most likely to appear in a phishing email?
    \begin{itemize}
        \item A. “Thank you for your registration! Click here to see your account details.”
        \item B. “Click here to view the latest collection of our awesome brand!”
        \item C. “Your bank account password has been compromised! If you don’t act fast, hackers might steal your money! Click here to reinitialize your password!” (Correct)
        \item D. “We noticed some unusual activity on your account. If this was not you, please contact us immediately.”
    \end{itemize}

    \item Which of the following messages is a phishing scam tactic?
    \begin{itemize}
        \item A. “This is a notification from the Cybercrime Division of your local police department... Pay a fine of \$500 within 24 hours to avoid prosecution.” (Correct)
        \item B. “We kindly request you to complete your account verification.”
        \item C. “You have won a vacation! Click here to claim.”
        \item D. “Please visit our website for new features.”
    \end{itemize}
\end{enumerate}

\subsection*{Phishing Warning Observation}

For each of 4 websites:

\begin{itemize}
    \item Carefully study the screenshot of the website
    \item Carefully read the warning generated by the security tool
    \item Briefly explain why the website was considered phishing
\end{itemize}

\subsection*{Website Assessment Task}

For each of 8 websites:

\begin{enumerate}
    \item Is the website Phishing or Benign?
    \item Explain what factors influenced your decision (e.g., URL, design, security indicators, spelling/grammar).
    \item How confident are you in your decision?
    \begin{itemize}
        \item Not at all confident
        \item Slightly confident
        \item Moderately confident
        \item Very confident
        \item Extremely confident
    \end{itemize}
\end{enumerate}

\subsection*{Post-Survey Questionnaire}

\begin{enumerate}
    \item How effective were the security warnings in guiding your decisions?
    \begin{itemize}
        \item Very helpful
        \item Somewhat helpful
        \item Neutral
        \item Not very helpful
        \item Not at all helpful
    \end{itemize}

    \item How has your confidence in identifying phishing websites changed?
    \begin{itemize}
        \item Much worse than before
        \item Slightly worse than before
        \item No change
        \item Slightly better than before
        \item Much better than before
    \end{itemize}

\end{enumerate}

\subsection{Example case-study of calculating Reliability of warning generation}. In Section~\ref{reliability}, we introduced our approach to calculate the reliability of the warnings generated by PhishXplain using four metrics: Correct Feature Rate (CFR), Feature Miss Rate (FMR) and Artifact Accuracy (AA) and Code Snippet Accuracy (CSA), with the final Reliability score calculated as: \noindent
Using these components, we compute the final \textbf{Reliability Score} (on a scale from 0 to 10) as:

\[
    \text{Reliability} = 10 \times \frac{\text{CFR} + (1 - \text{FMR}) + \text{AA} + \text{CSA}}{4}
\]

\noindent

Here we show an example of this calculation. Lets say PhishXplain returns the following features: (1) \textit{Spelling errors and typos}, (2) \textit{Requests for sensitive data}, (3) \textit{Threats or penalties}, and (4) \textit{Fake countdown timers}. 
Upon manual inspection, the coders confirmed that the first three features were indeed present on the website, while the fourth was not. 
For the three valid features, two had correctly inserted artifacts grounded in the source code (e.g., actual typos and threatening messages), while one contained a hallucinated artifact.
Similarly, two out of the four predicted features had accurately identified the code snippets necessary for drawing bounding boxes on the rendered warning screenshot.

The coders independently annotated the ground truth for this website, which contained exactly three phishing features: spelling errors, a sensitive data request, and a threatening message. 
Using this reference, we compute the evaluation metrics as follows: the Correct Feature Rate (CFR) is $3/4 = 0.75$, since three out of four predicted features were valid. The Feature Miss Rate (FMR) is $0$, as the model identified all ground-truth features; thus, $(1 - \text{FMR}) = 1.0$. 
Artifact Accuracy (AA) is $2/3 \approx 0.67$, based on two accurate artifacts out of three valid features. Code Snippet Accuracy (CSA) is $2/4 = 0.5$ since only two predictions had correct HTML tag localization. 
The final reliability score is calculated as the mean of the four sub-scores, scaled to a 10-point scale:

\[
\text{Reliability} = 10 \times \frac{0.75 + 1.0 + 0.67 + 0.5}{4} = 10 \times \frac{2.92}{4} = 7.3
\]

\section*{Websites used in User-study}
This section illustrates the 12 websites that were shown to the participants. Besides the 4 websites that were accompanied with warnings, the other eight websites were shown in random order to avoid recency bias.
\begin{figure}[H]
\centerline{\includegraphics[width=0.7\columnwidth]{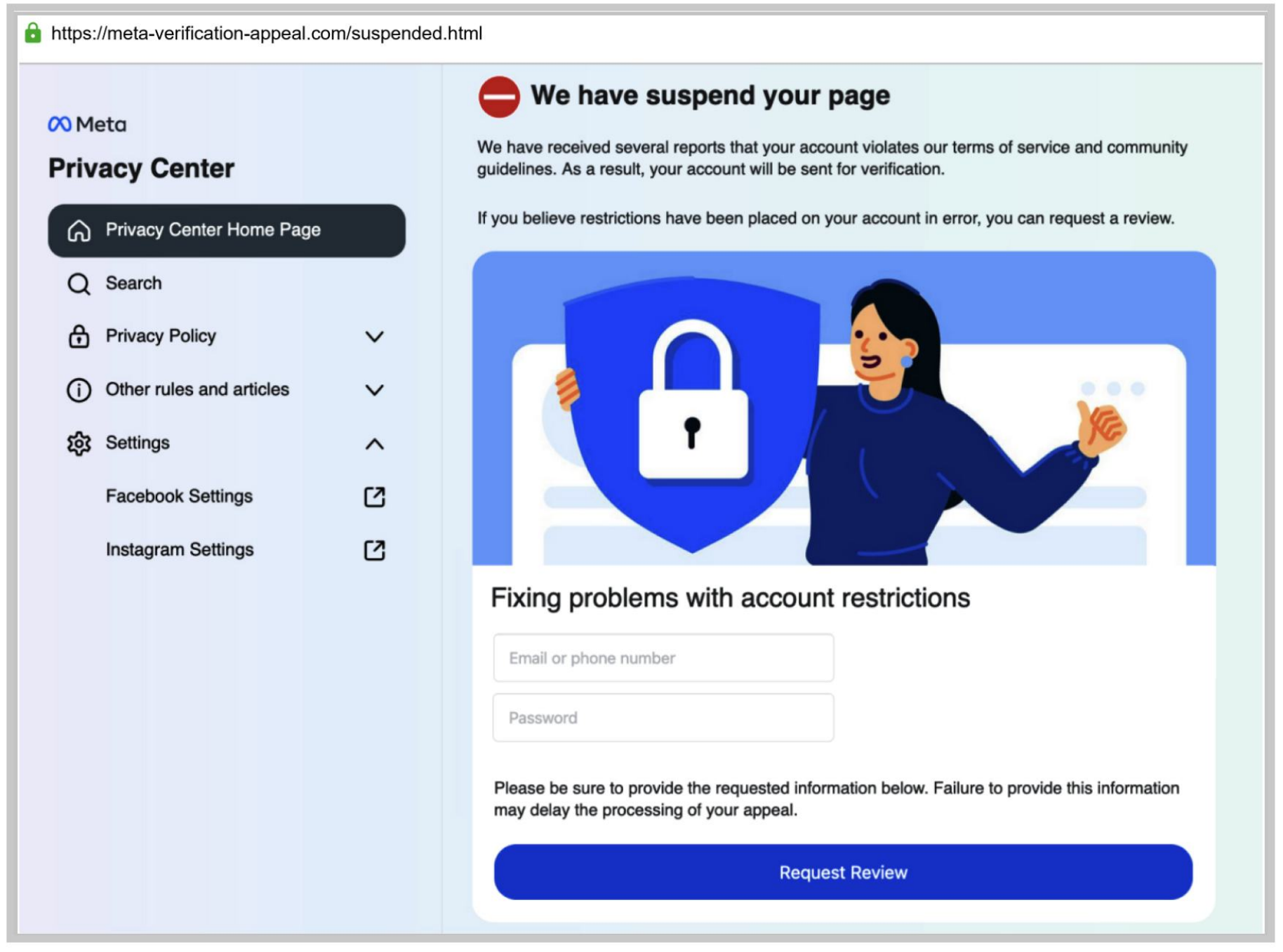}}
\caption*{Warning-1}
\label{fig:w1}
\end{figure}

\begin{figure}[H]
\centerline{\includegraphics[width=0.7\columnwidth]{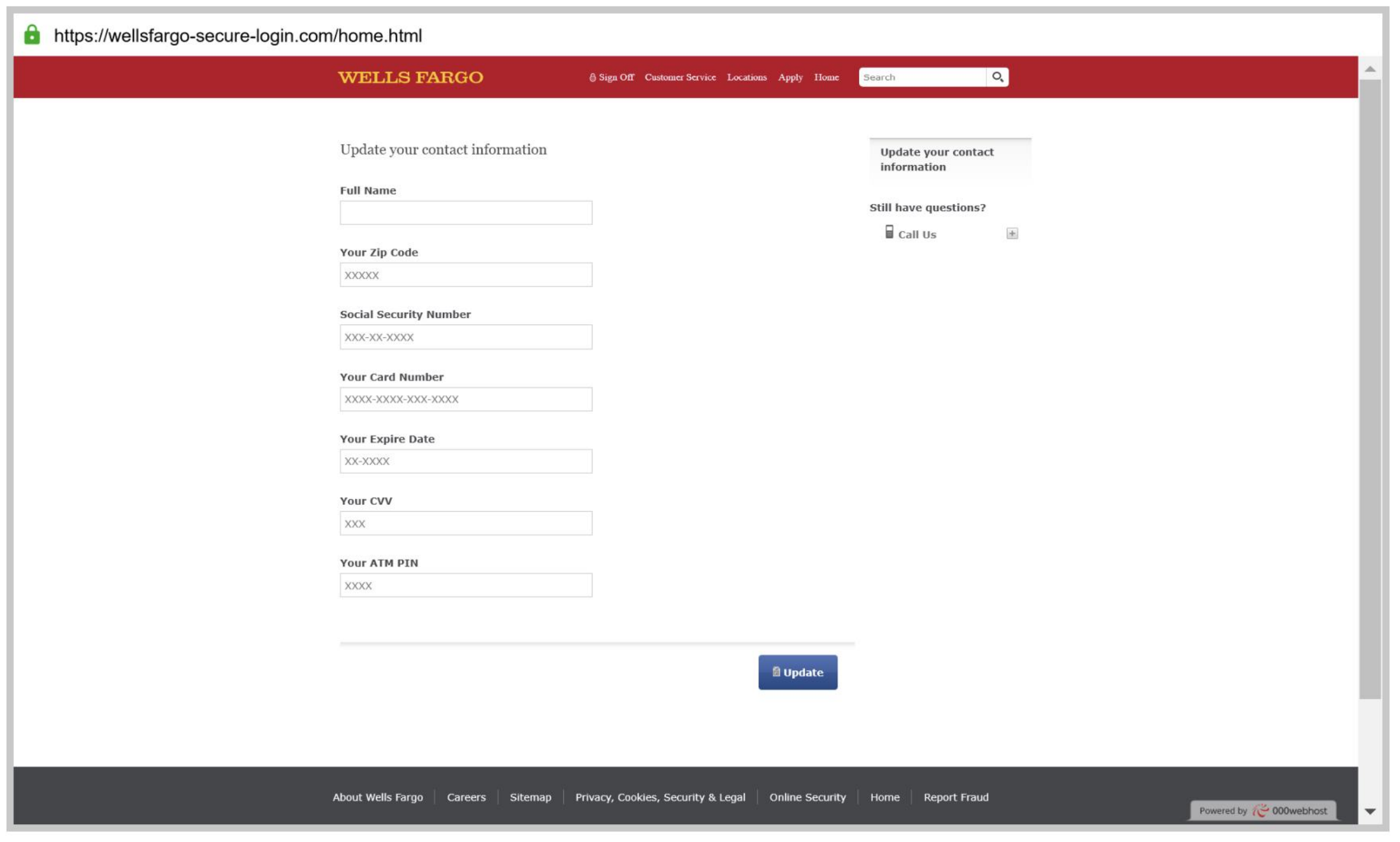}}
\caption{Warning-2}
\label{fig:w2}
\end{figure}

\begin{figure}[H]
\centerline{\includegraphics[width=0.7\columnwidth]{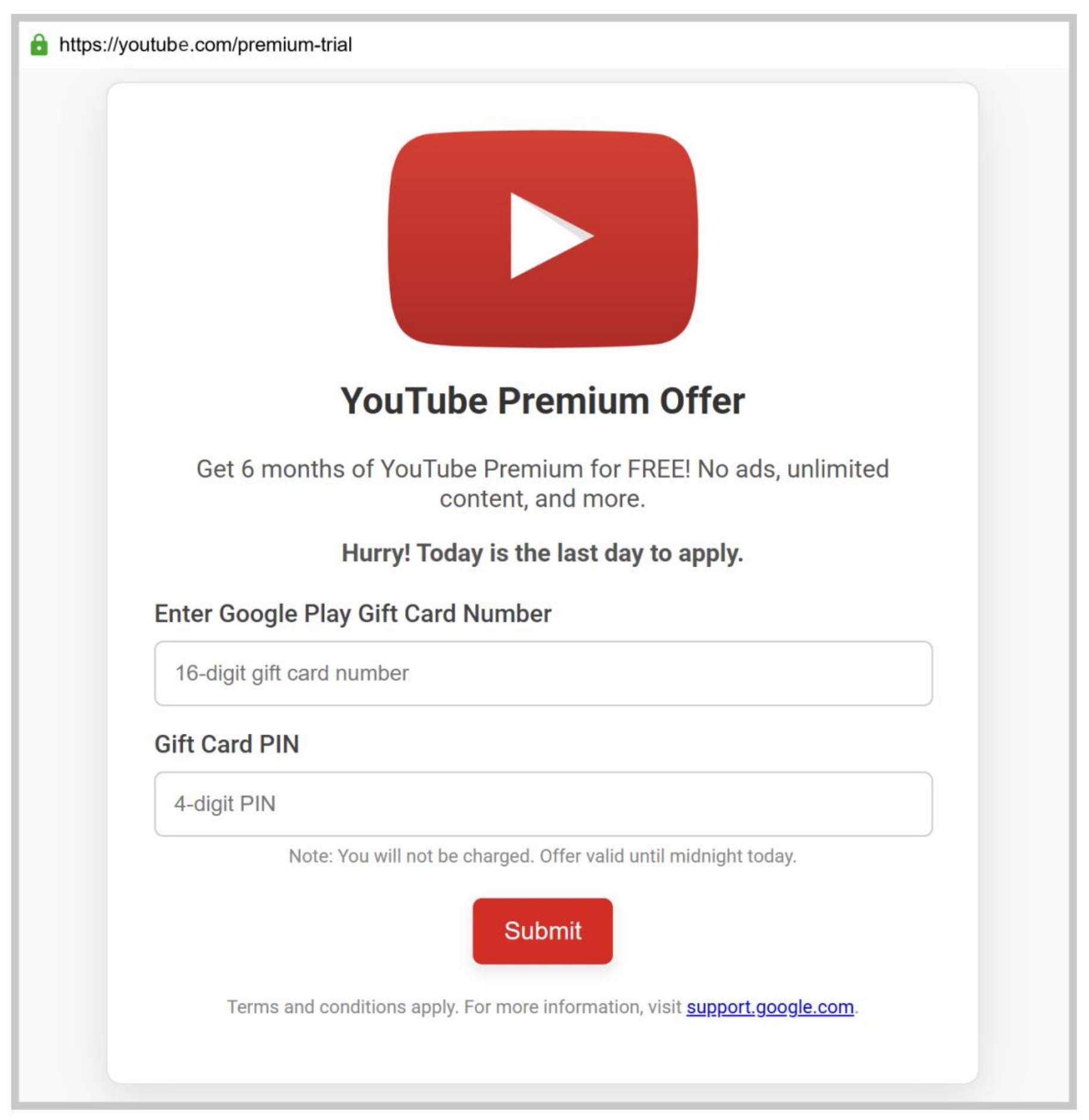}}
\caption{Warning-3}
\label{fig:w3}
\end{figure}

\begin{figure}[H]
\centerline{\includegraphics[width=0.7\columnwidth]{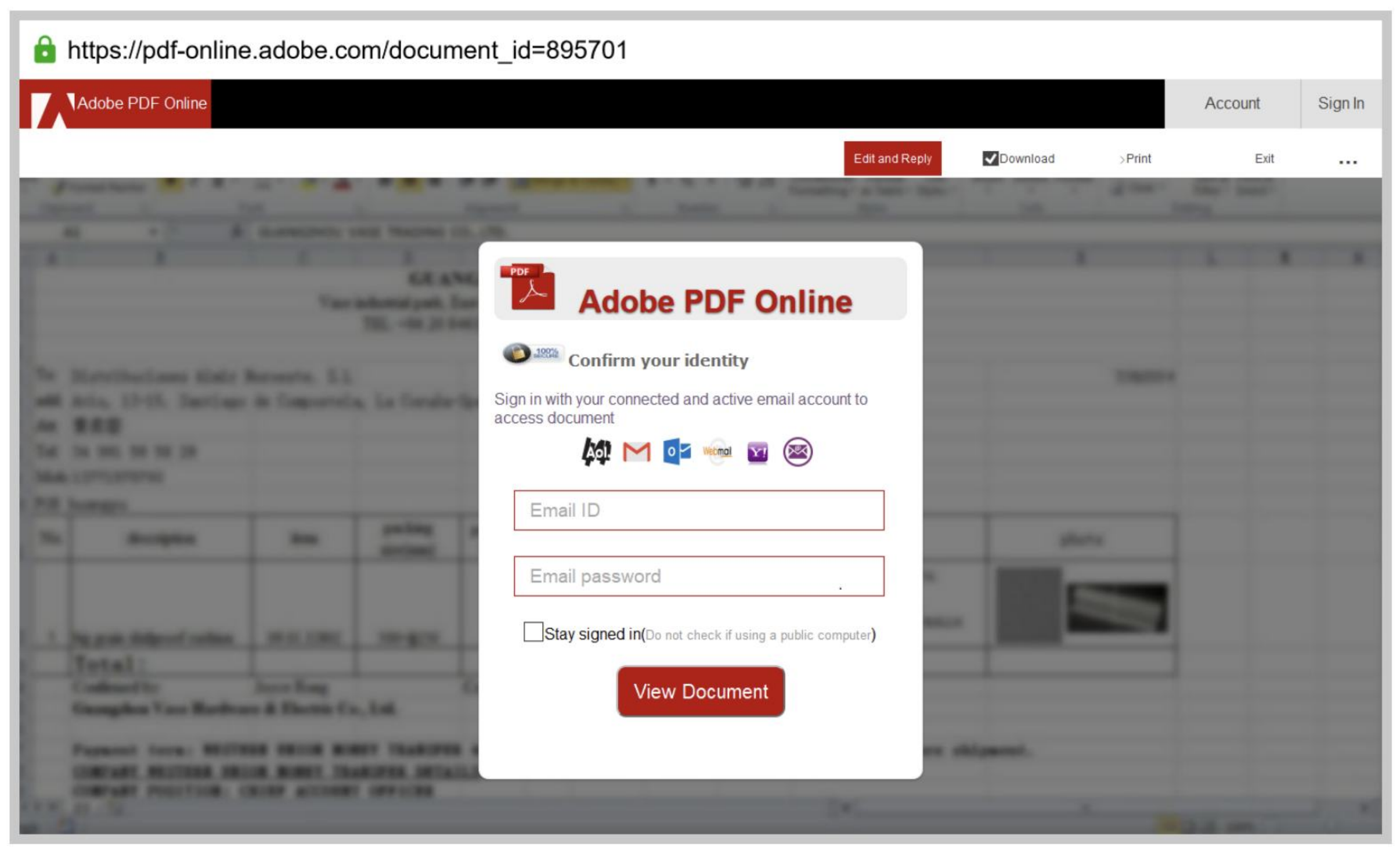}}
\caption*{Warning-4}
\label{fig:w4}
\end{figure}

\begin{figure}[H]
\centerline{\includegraphics[width=0.7\columnwidth]{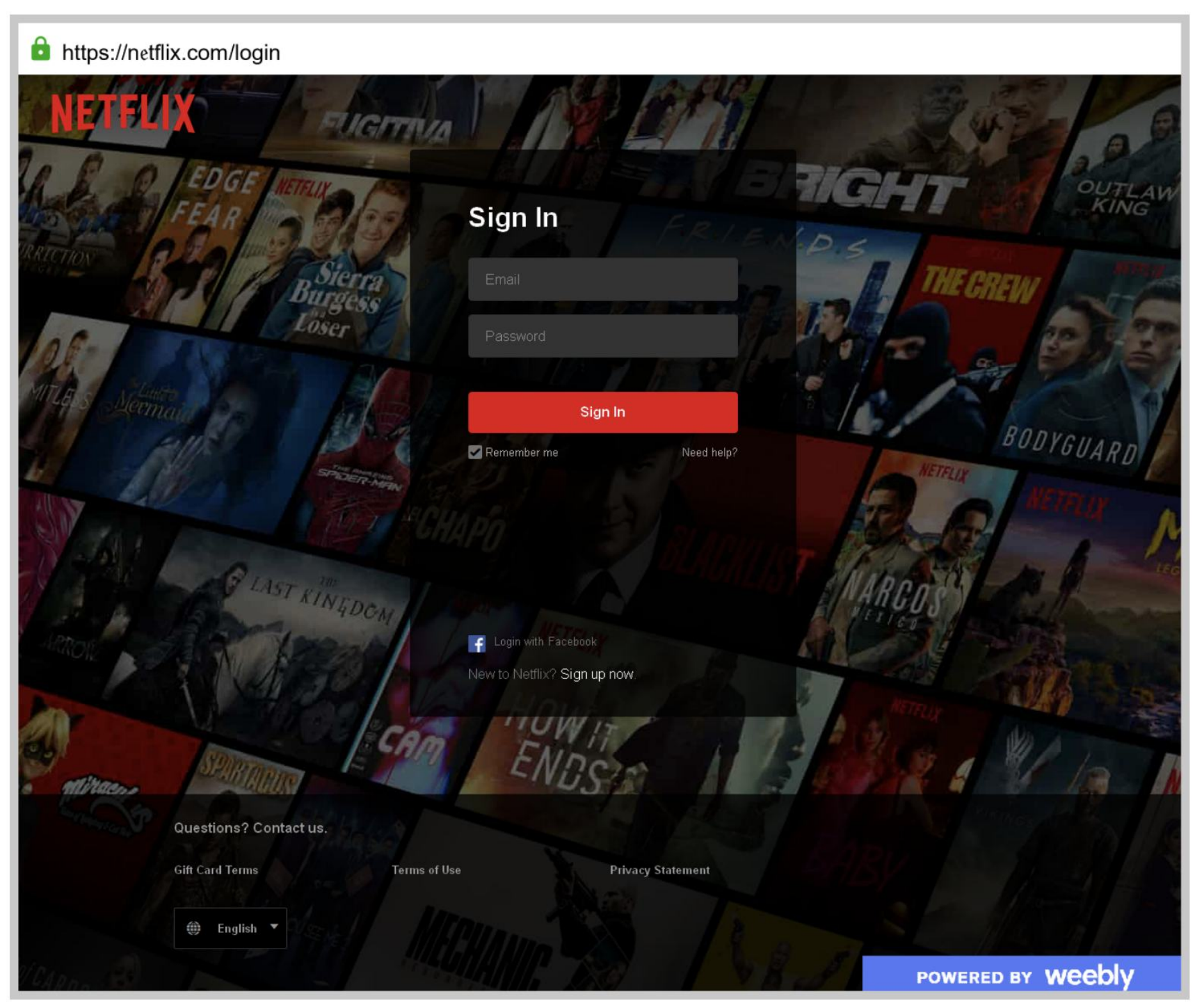}}
\caption*{Phishing-1}
\label{fig:p1}
\end{figure}

\begin{figure}[H]
\centerline{\includegraphics[width=0.7\columnwidth]{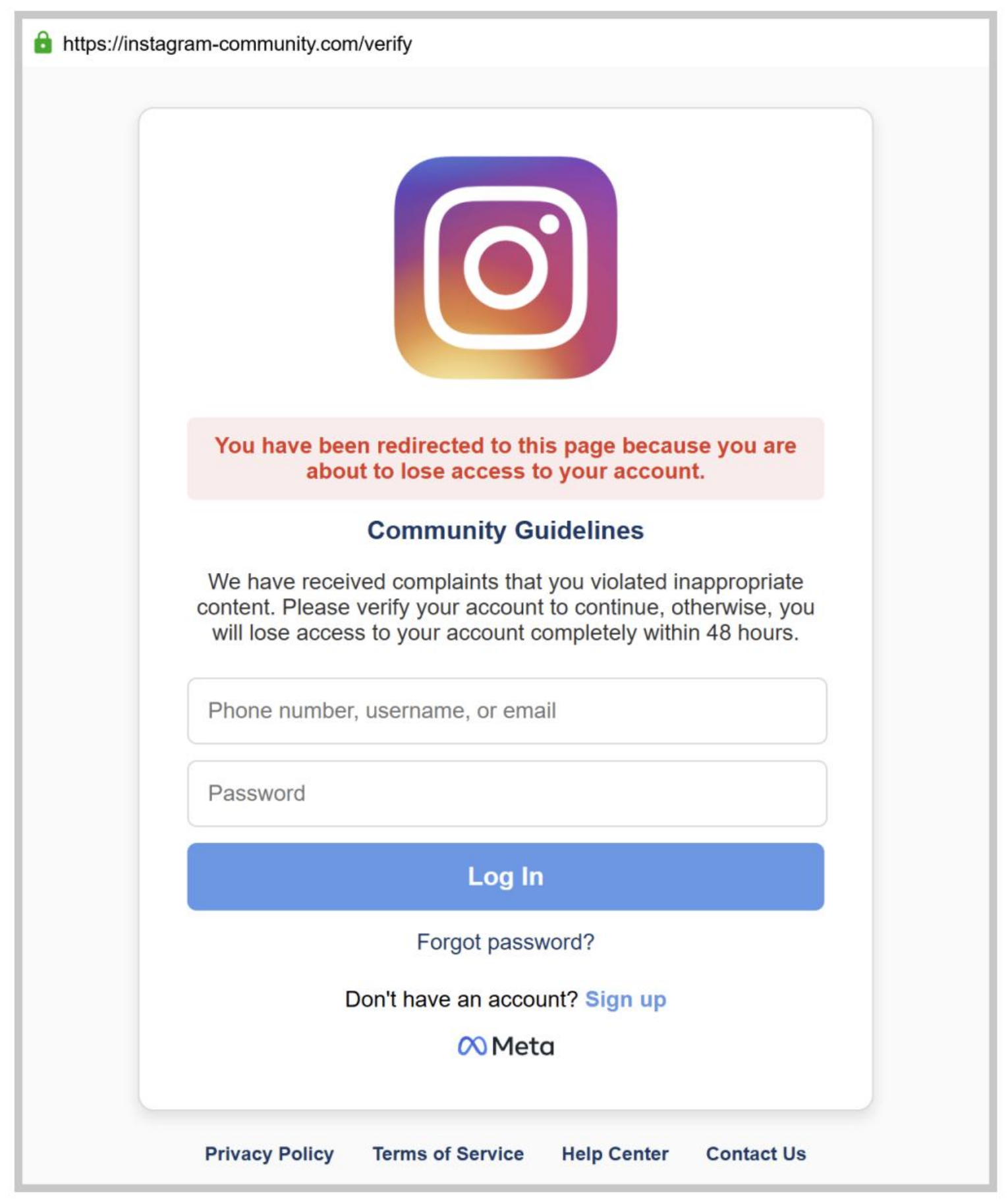}}
\caption*{Phishing-2}
\label{fig:p2}
\end{figure}

\begin{figure}[H]
\centerline{\includegraphics[width=0.7\columnwidth]{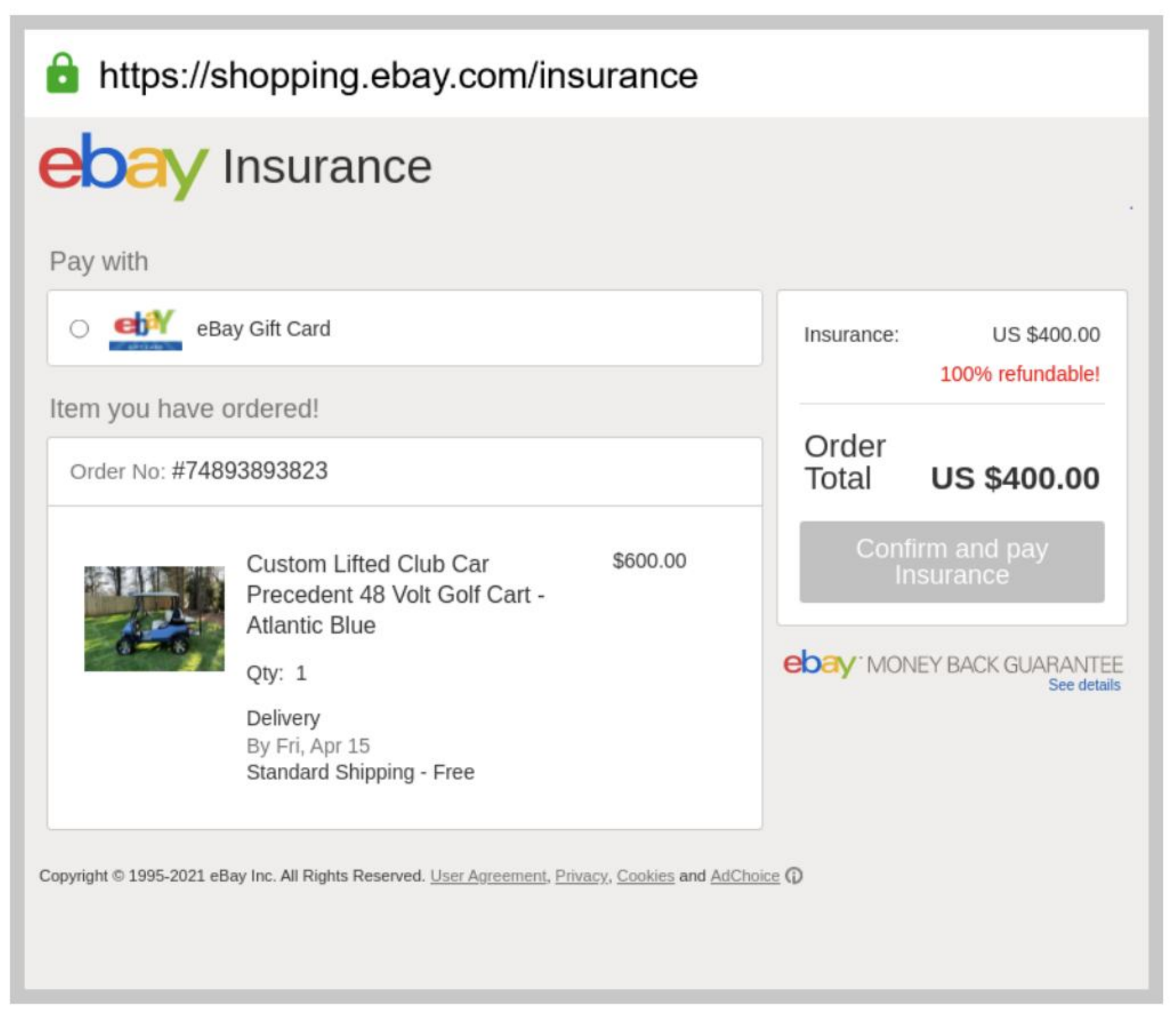}}
\caption*{Phishing-3}
\label{fig:p3}
\end{figure}

\begin{figure}[H]
\centerline{\includegraphics[width=0.7\columnwidth]{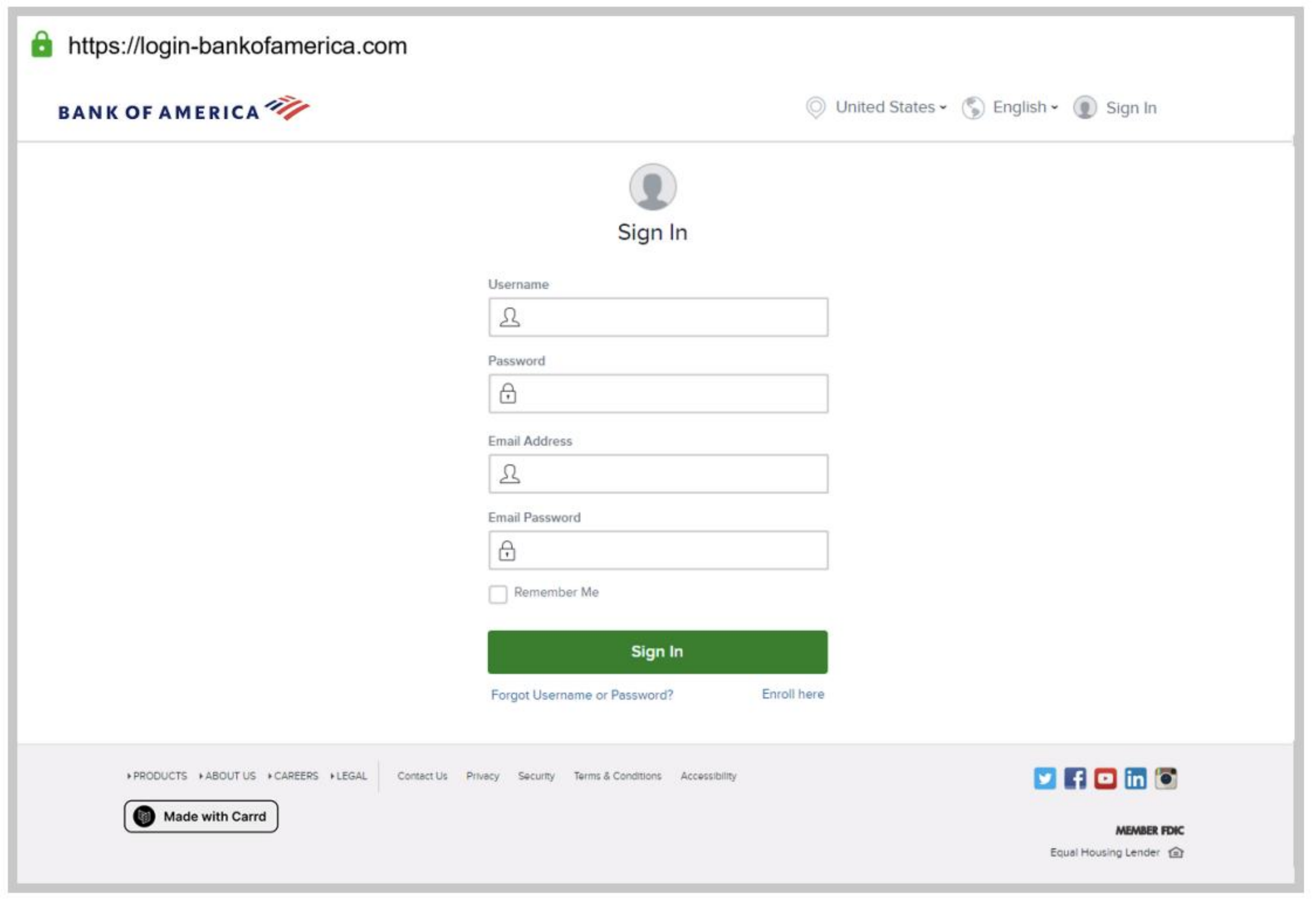}}
\caption*{Phishing-4}
\label{fig:p4}
\end{figure}

\begin{figure}[H]
\centerline{\includegraphics[width=0.7\columnwidth]{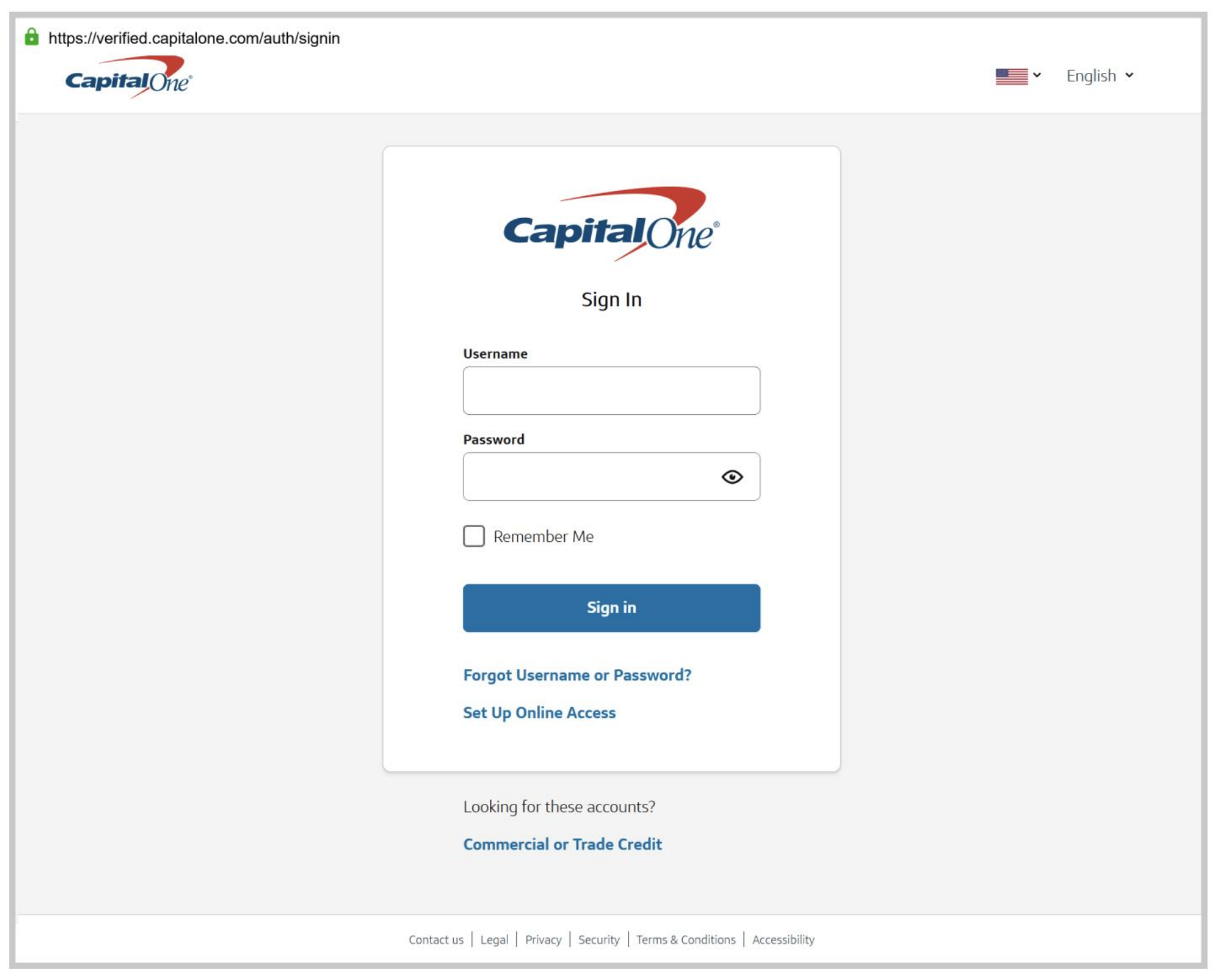}}
\caption*{Benign-1}
\label{fig:b1}
\end{figure}

\begin{figure}[H]
\centerline{\includegraphics[width=0.7\columnwidth]{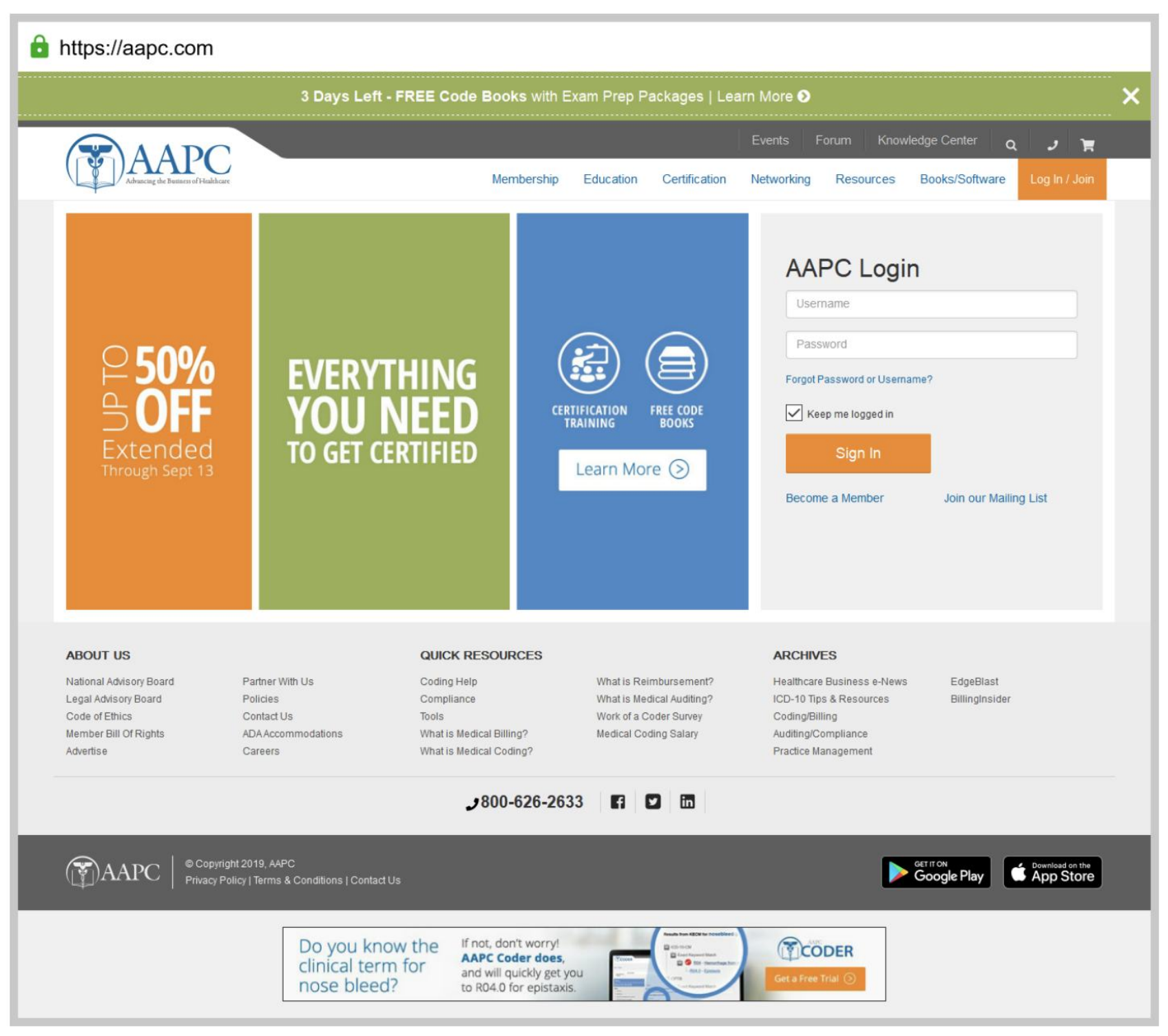}}
\caption*{Benign-2}
\label{fig:b2}
\end{figure}

\begin{figure}[H]
\centerline{\includegraphics[width=0.7\columnwidth]{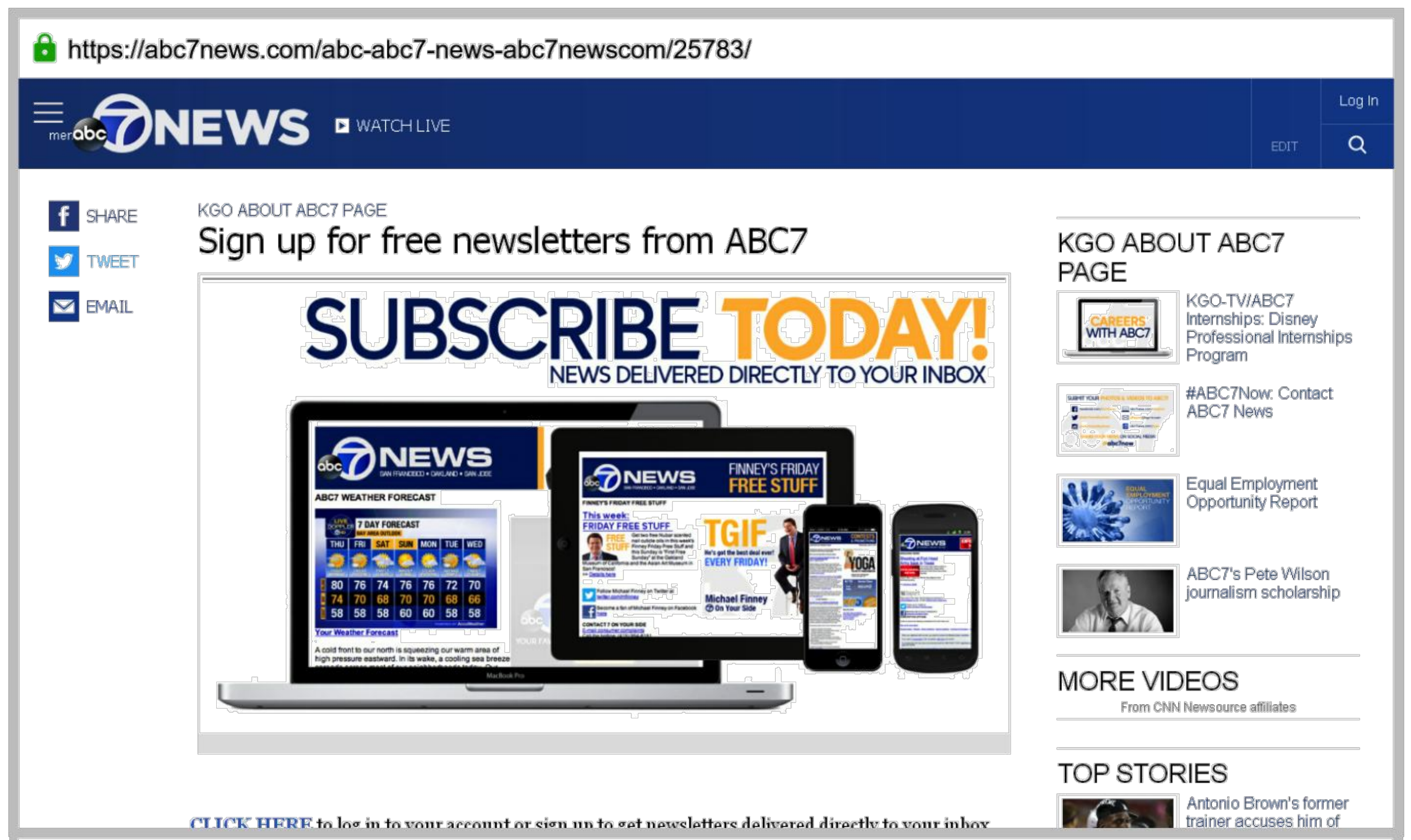}}
\caption*{Benign-3}
\label{fig:b3}
\end{figure}

\begin{figure}[H]
\centerline{\includegraphics[width=0.7\columnwidth]{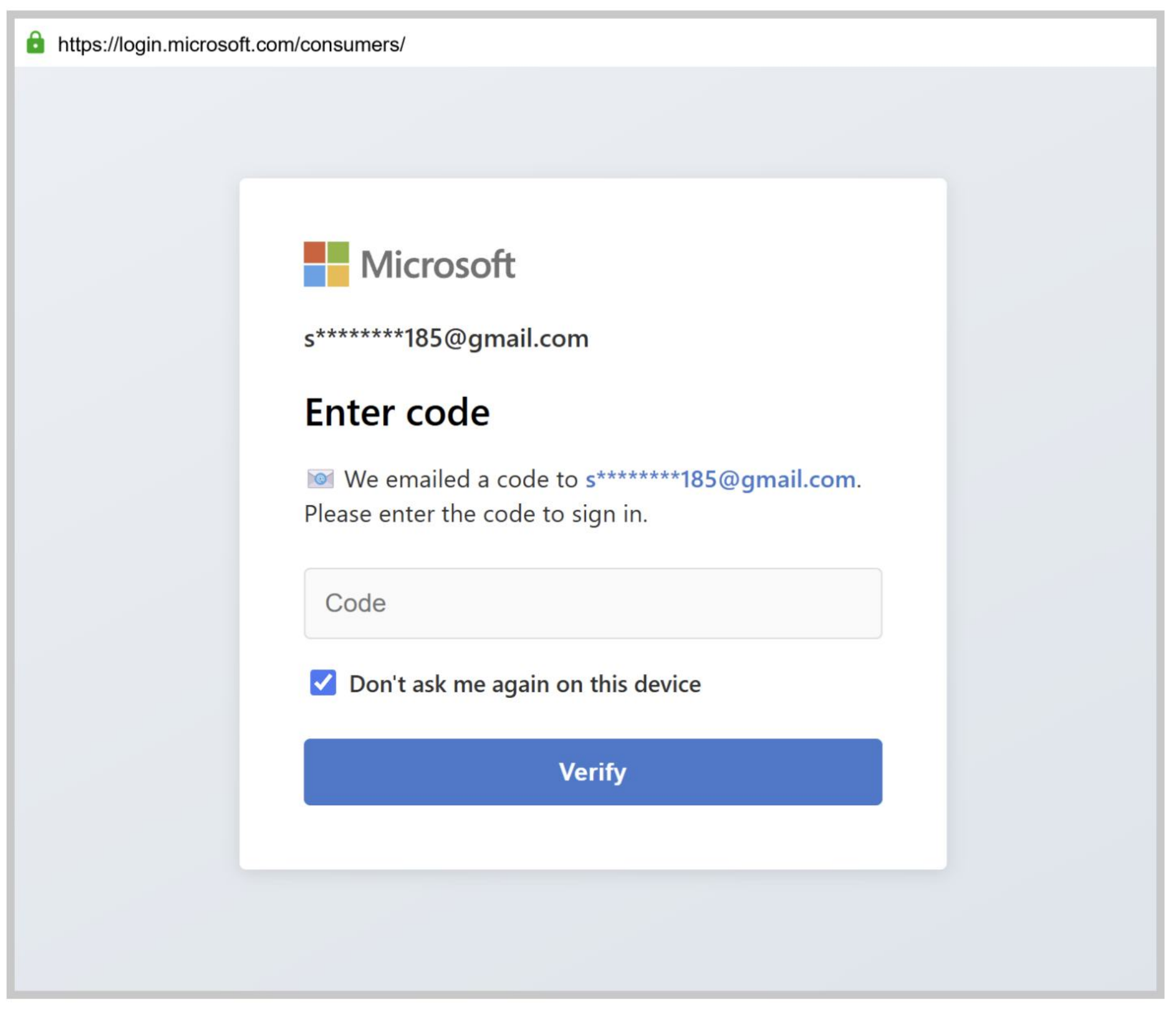}}
\caption*{Benign-4}
\label{fig:b4}
\end{figure}

\subsection*{Assorted components}

\begin{table}[H]
\caption{Participants' Demographic characteristics by group}
\centering
\resizebox{0.65\columnwidth}{!}{%
\begin{tabular}{lcc}
\hline
\textbf{Variable} & \textbf{Group A} & \textbf{Group B} \\
\hline
\textbf{Gender} & & \\
Male & 29 & 25 \\
Female & 43 & 46 \\
Non-binary & 3 & 3 \\
\hline
\textbf{Age} & & \\
18--24 & 10 & 7 \\
25--34 & 30 & 21 \\
35--44 & 12 & 22 \\
45--54 & 12 & 13 \\
55+ & 11 & 11 \\
\hline
\textbf{Education Level} & & \\
High School or Equivalent & 3 & 10 \\
Some College & 24 & 22 \\
Bachelor's Degree & 36 & 32 \\
Master's Degree & 10 & 7 \\
Doctorate & 2 & 3 \\
\hline
\end{tabular}}
\label{tab:demographics}
\end{table}

\begin{figure}[H]
\centerline{\includegraphics[width=1\columnwidth]{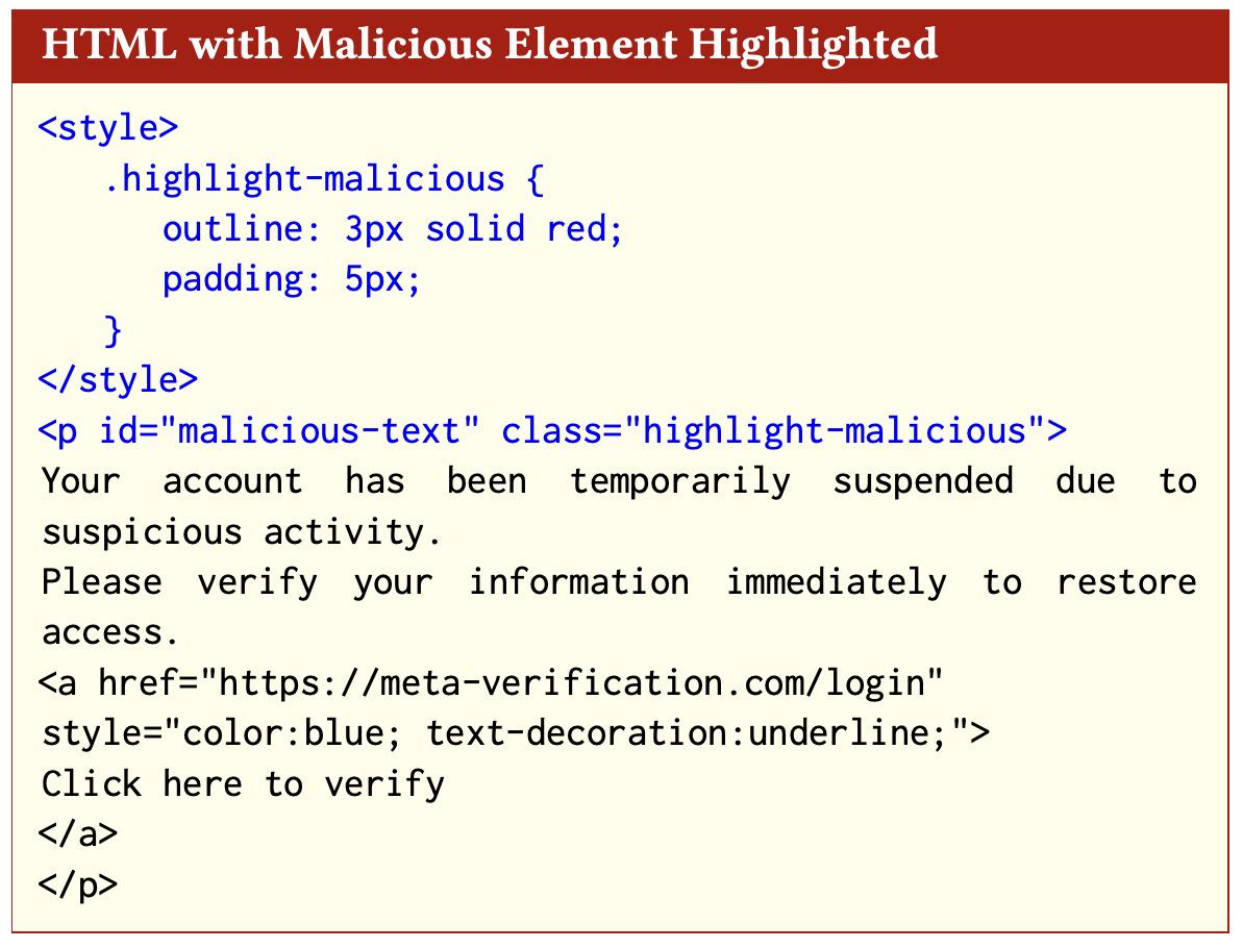}}
\caption*{Code-block where the malicious feature is highlighted}
\label{fig:code_block}
\end{figure}

\end{document}